\pdfoutput=1
\documentclass[aps,twocolumn,superscriptaddress]{revtex4-1}
\usepackage{amsmath}
\usepackage{amssymb}
\usepackage{amsfonts}
\usepackage[overload]{empheq}
\usepackage{graphicx} %package to manage images
\graphicspath{ {./Figs/} }
\usepackage[margin=0.75in]{geometry}
\usepackage[utf8]{inputenc}
\usepackage[english]{babel}
\usepackage[colorinlistoftodos]{todonotes}
\usepackage{mathtools}
\usepackage{float}
\usepackage{array,multirow}
\usepackage{bm}
\usepackage{cases}
\usepackage[position=top]{subfig}
\usepackage{pdfpages}

\makeatletter
\AtBeginDocument{\let\LS@rot\@undefined}
\makeatother

\newcommand{\meff}{M_{\text{eff}}}

\newcommand{\xidl}{\xi_{\text{DL}}}
\newcommand{\xifl}{\xi_{\text{FL}}}
\newcommand{\xifmr}{\xi_{\text{FMR}}}

\newcommand{\NM}{\hat{m}}
\newcommand{\tip}{\tau_x}
\newcommand{\toop}{\tau_z}
\newcommand{\tfm}{t_\text{FM}}
\newcommand{\tnm}{t_\text{HM}}
\newcommand{\geff}{g^{\uparrow\downarrow}_\text{eff}}

\newcommand{\lsd}{\lambda_\text{sd}}

\renewcommand{\Re}[1]{\text{Re}\left[#1\right]}
\newcommand{\Ramr}{R_\text{AMR}}
\newcommand{\Rahe}{R_\text{AHE}}
\newcommand{\Rphe}{R_\text{PHE}}
\newcommand{\Irf}{I_\text{RF}}

\begin{document}

\title{Transverse and Longitudinal Spin-Torque Ferromagnetic Resonance for Improved Measurements of Spin-Orbit Torques}
\author{Saba Karimeddiny}
    %\email[Correspondence email address: ]{email@institution.com}% Your name
    \thanks{These authors contributed equally}
    \affiliation{Cornell University, Ithaca, NY 14850, USA}
\author{Joseph A. Mittelstaedt}
    %\email[Correspondence email address: ]{email@institution.com}% Your name
    \thanks{These authors contributed equally}
    \affiliation{Cornell University, Ithaca, NY 14850, USA}
\author{Robert A. Buhrman}
    %\email[Correspondence email address: ]{email@institution.com}% Your name
    \affiliation{Cornell University, Ithaca, NY 14850, USA}
\author{Daniel C. Ralph}
    %\email[Correspondence email address: ]{email@institution.com}% Your name
    \affiliation{Cornell University, Ithaca, NY 14850, USA}
    \affiliation{Kavli Institute at Cornell, Ithaca, NY 14853, USA}

\date{\today} % Leave empty to omit a date

\begin{abstract}
Spin-torque ferromagnetic resonance (ST-FMR) is a common method used to measure spin-orbit torques (SOTs) in heavy metal/ferromagnet bilayer structures.  In the course of a measurement, other resonant processes such as spin pumping (SP) and heating can cause spin current or heat flows between the layers, inducing additional resonant voltage signals via the inverse spin Hall effect (ISHE) and Nernst effects (NE). In the standard ST-FMR geometry, these extra artifacts exhibit a dependence on the angle of an in-plane magnetic field that is identical to the rectification signal from the SOTs.  We show experimentally that the rectification and artifact voltages can be quantified separately by measuring the ST-FMR signal transverse to the applied current (i.e., in a Hall geometry) in addition to the usual longitudinal geometry. We find that in Pt (6 nm)/CoFeB samples the contribution from the artifacts is small compared to the SOT rectification signal for CoFeB layers thinner than 6 nm, but can be significant for thicker magnetic layers.  We observe a sign change in the artifact voltage as a function of CoFeB thickness that we suggest may be due to a competition between a resonant heating effect and the SP/ISHE contribution.    
\end{abstract}

\maketitle

\section{Introduction}
Current-induced spin-orbit torques (SOTs) have the potential to provide improved efficiency in the control of magnetic memory and logic devices, enabling new technologies that are fast, non-volatile, high-density, and of infinite endurance \cite{Brataas2012,Wang2013,Oboril2015}.  
The metrology of SOT materials and  devices is critical to these developments.  Several different techniques have been developed to quantify spin-orbit torques, including spin-torque ferromagnetic resonance (ST-FMR) \cite{Liu2011,Liu2012,Mellnik2014}, second-harmonic (low-frequency) Hall measurements \cite{Pi2010,Garello2013,Hayashi2014}, optical measurements of current-induced magnetic deflection \cite{Fan2014,Fan2016}, determination of the threshold currents for switching of nanoscale magnets with in-plane anisotropy \cite{Sun2000, Liu2012}, measurements of spin Hall magnetoresistance \cite{Althammer2013,Kim2016}, and measurements of current-induced domain wall motion within perpendicular magnetic films \cite{Emori2014,Pai2016}. However, different techniques sometimes produce inconsistent results \cite{Pai2015, Tao2018} and can even give internal discrepancies. For example, independent second harmonic Hall studies on layers with in-plane and out-of-plane magnetic anisotropy \cite{Zhu2019AFM,Lau2017} have measured discrepant (and sometimes unphysical) results for the damping-like torque efficiency $\xidl$, and ST-FMR and second-harmonic Hall measurements on samples with in-plane anisotropy can differ by tens of percent.  Therefore, there is a continuing need to examine possible artifacts affecting the different measurement approaches and to improve their accuracy.  
\par
Here we consider one of the most popular techniques to measure SOTs, ST-FMR.  A known artifact in ST-FMR is that the measured signals can include contributions from spin pumping (SP) together with the inverse spin Hall effect (ISHE) \cite{Tserkovnyak2002,Tserkovnyak2002a,Mosendz2010a,Azevedo2011}. In addition, there can be  thermoelectric contributions resulting from resonant heating that gives rise to a longitudinal spin Seebeck effect (LSSE) together with the ISHE \cite{Uchida2010,Holanda2017}, or Nernst effects (NE) \cite{Lee2015, kikkawa2013, Avci2014, Roschewsky2019}. In the standard ST-FMR measurement configuration, these artifact signals are challenging to disentangle from the primary spin-torque diode (rectification) signal because they all have identical dependences on the angle of a magnetic field applied within the device plane \cite{Mosendz2010a, Avci2014}.  
\par
Previous studies attempting to separate artifact voltages from the ST-FMR signal have largely been focused on SP/ISHE contributions \cite{Kondou2016, Okada2019, Kumar2019}. One previous study has attempted to separate SP/ISHE by using the external field to tilt the magnetization partly out of plane \cite{Okada2019}, but this configuration can be tricky to implement and interpret due to the large demagnetization fields of typical devices and the possibility of spatially non-uniform magnetization states.
We demonstrate a straightforward alternative approach to separately quantify both the spin-orbit torque and the spin-pumping/resonant-heating artifact signals using only in-plane magnetic fields, by measuring the ST-FMR signal transverse to the applied current (i.e., in a Hall geometry) in addition to the usual longitudinal geometry.

\section{Background}

In conventional ST-FMR, a microwave current is injected along a rectangular sample of a heavy metal (HM)/ferromagnet (FM) bilayer to induce FMR through current-induced torques acting on the magnetization. Within a simple macrospin model, the Landau-Lifshitz-Gilbert-Slonczewski (LLGS) equation captures the resulting dynamics of the magnetic moment:
\begin{align}
\begin{split}
    \dot{\NM} &= \gamma \NM\times\frac{dF}{d\NM} \; + \; \alpha \NM\times \dot{\NM} \\ &\quad\quad + \tau^0_\text{DL}\NM\times\left(\hat{\sigma}\times\NM\right) 
     \;+ \;  \tau^0_\text{FL}\hat{\sigma}\times\NM
    \label{LLGS}
\end{split}
\end{align}
where $\NM$ is the normalized magnetic moment of the FM, $F$ is the free energy density of the FM, $\gamma=2\mu_B/\hbar$ is the gyromagnetic ratio with $\mu_B$ the Bohr magneton, and $\alpha$ is the Gilbert damping parameter. The final two terms represent the current-induced damping-like and field-like torques, with prefactors
\begin{equation}\label{taudl}
     \tau^0_\text{DL} = \xi_\text{DL}\frac{\mu_B J_e}{e M_S t_\text{FM}}
\end{equation}
\begin{equation}\label{taufl}
    \tau^0_\text{FL} = \xi_\text{FL}\frac{\mu_B J_e}{e M_S t_\text{FM}}.
\end{equation}
Here $\xi_\text{DL}$ and $\xi_\text{FL}$ are dimensionless spin-torque efficiencies that one might wish to measure for a given material system.  $J_e$ is the charge current density in the HM, $e$ is the magnitude of the electron charge, $M_S$ is the saturation magnetization of the FM, $t_\text{FM}$ is the thickness of the ferromagnetic layer, and $\hat{\sigma}$ denotes the polarization of the spin current incident on the ferromagnet.     For a non-magnetic heavy metal with an ordinary high-symmetry crystal structure, $\hat{\sigma}$ is required by symmetry to be in-plane and perpendicular to the applied current so that, for an in-plane magnetization, the damping-like torque points in the sample plane and the field-like torque points out of plane; we will assume this to be the case throughout this paper.
\par
The magnetic resonance can be detected via a rectified longitudinal DC voltage (oriented along the length of the wire parallel to the current) caused by the mixing of the microwave current with resistance oscillations produced by the precessing magnet via the anisotropic magnetoresistance (AMR) or spin Hall magnetoresistance (SMR) \cite{Tulapurkar2005,Sankey2006}. The resonance peak shape as a function of magnetic field magnitude at a constant field angle for this rectified signal is the sum of symmetric and antisymmetric Lorentzian functions. For a magnetic layer with in-plane anisotropy and and in-plane magnetic field, the symmetric component arises from $\tau^0_\text{DL}$ and the antisymmetric component from the combination of the current-induced Oersted field and $\tau^0_\text{FL}$.  Once the microwave current is calibrated, the measurement allows determinations of both $\xi_\text{DL}$ and $\xi_\text{FL}$, assuming there are no other artifacts contaminating the signal. 
\par
When the FM layer is resonantly excited, a pure spin current resulting from SP or LSSE can also flow from the FM layer into the HM layer and produce a measurable voltage through the ISHE of the HM \cite{Tserkovnyak2002,Tserkovnyak2002a,Mosendz2010a,Azevedo2011, Uchida2010, Lustikova2015, jungfleisch2015, Nakayama2012, Rezende2014}. Furthermore, an out-of-plane temperature gradient within the heterostructure due to resonant heating can produce a thermoelectric voltage from ordinary or anomalous Nernst effects \cite{Holanda2017,Roschewsky2019}. In all of these processes, the result is a DC voltage perpendicular to the magnetization axis with a symmetric Lorentzian lineshape \cite{Saitoh2006, Iguchi2017, Mosendz2010a}. Consequently, if these artifact signals are sufficiently large, they can contaminate ST-FMR measurements of $\tau^0_\text{DL}$.  The signals from spin-torque rectification and the spin-pumping/resonant-heating artifacts all have the same dependence on the angle of an in-plane magnetic field: $\propto \sin(2\phi)\cos(\phi)$, with $\phi$ measured relative to the positive applied current direction  \cite{Mosendz2010a, Kondou2016, Kumar2019, Avci2014}, making artifact effects difficult to disentangle.
\par
In this work, we demonstrate that if one performs a ST-FMR experiment as a function of the angle of an in-plane magnetic field by measuring the resonant DC voltage {\em transverse} to the current (i.e., in a Hall geometry) the rectified spin-torque contribution and the spin pumping/resonant heating can be distinguished. We are aware of previous works that have performed ST-FMR in the transverse geometry \cite{bose2017,Kumar2019}, but these studies did not illustrate how to separate the rectified spin-torque contribution from the artifact signals. A closely-related idea was used previously in experiments which studied SP/ISHE signals from magnetic precession excited using oscillating magnetic fields, in order to separate out unwanted (in that context) rectification signals \cite{Lustikova2015, Keller2017}.  Harder et al.\ have published a review mapping out the field-angle dependence expected for resonance experiments in both longitudinal and transverse geometries for different orientations of excitation \cite{harder2016}.
\par

\section{Theory}
\begin{figure}[t]
\begin{tabular}{cc}
\subfloat[]{\includegraphics[width = 0.48\linewidth]{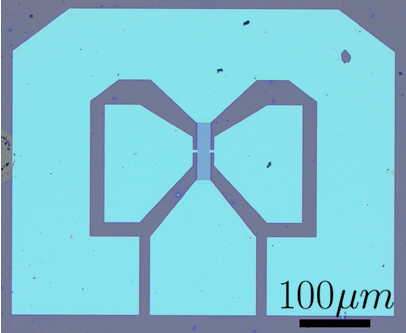}}&
\subfloat[]{\includegraphics[width = 0.48\linewidth]{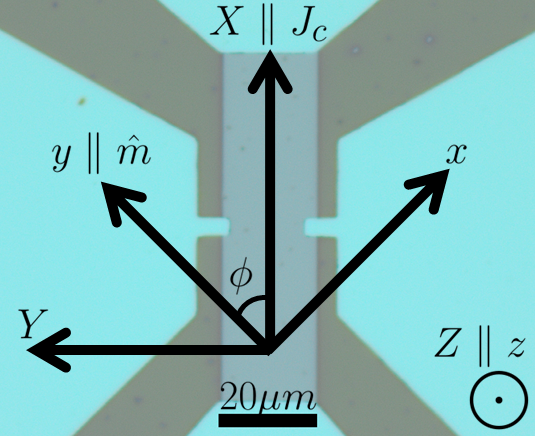}}
\end{tabular}
\caption{\textbf{(a)} Optical image of our Hall ST-FMR device, showing the geometry of the contact pads. This particular device featured a Pt(6)/CoFeB(6) bilayer measuring 20$\times$80 $\mu$m$^2$ (in the center, dark blue). The scale bar is 100 $\mu$m. \textbf{(b)} Zoomed-in optical image of the bilayer and contacts with our coordinate definitions. The $XYZ$ (capital) coordinates are fixed relative to the device geometry while $xyz$ (lowercase) coordinates are relative to the equilibrium orientation of the magnetization. The scale bar is 20 $\mu$m.}
\label{fig:1}
\end{figure}

We consider a thin-film macrospin magnet with in-plane anisotropy subject to an external in-plane magnetic field oriented at an angle $\phi$ with respect to the positive current direction, that aligns the equilibrium direction of the magnetization (see Fig.\ \ref{fig:1}).  We define the $\hat{y}$ axis to be parallel to the equilibrium direction of the magnetization and $\hat{z}$ to be perpendicular to the sample plane so that $\hat{x} =\hat{y}\times\hat{z}$ is in-plane.  We will also use capital letters to indicate a separate coordinate system fixed with respect to the sample, where $\hat{X}$ is along the current direction, $\hat{Z}=\hat{z}$, and $\hat{Y}=\hat{Z} \times \hat{X}$. Spherical polar coordinates $\theta,\phi$ for the magnetization orientation are defined relative to the $X,Y,Z$ axes.
\\
\par
A microwave current $I_\text{RF} \Re{e^{-i\omega t}}$ is applied, producing alternating torques with amplitudes $\tau_x = \tau^0_\text{DL}\cos(\phi)$  and $\tau_z = \tau^0_z \cos(\phi)= (\tau^0_\text{FL}+\tau^0_\text{Oe})\cos(\phi)$ in the $\hat{x}$ and $\hat{z}$ directions.  With these definitions,  $\tau^0_\text{Oe}$ takes a positive value by Ampere's Law and $\tau^0_\text{DL}$ is positive for the spin Hall effect of Pt.  Linearization and solution of the LLGS equation (see Supplmentary Information \cite{Supplement}) allows us to calculate the oscillatory components of the magnetic moment, in complex notation,
\begin{align}\label{mymz}
\begin{split}
    	m_x &= \frac{-\omega_2\toop + i\omega\tip}{-\gamma(B - B_0)\omega^+ + i\alpha\omega\omega^+}\\
    	m_z &= \frac{\omega_1\tip + i\omega\toop}{-\gamma(B - B_0)\omega^+ + i\alpha\omega\omega^+}.
\end{split}
\end{align}
Here $B_0$ is the resonance field, $B$ is the applied external field, $\omega_1 = \gamma B_0$, $\omega_2 = \gamma(B_0 + \mu_0\meff)$, and $\omega^+ = \omega_1 + \omega_2$; $\meff$ is the in-plane saturation magnetization ($M_S$) minus any out-of-plane anisotropy. Note that by our definition of coordinate axes, during the precession $m_x = - d\phi$ and $m_z = -d\theta$.

Assuming that the anisotropic magnetoresistance has the form 
$R_{XX} = R_0 + R_{\text{AMR}}m_X^2$,
the spin-torque mixing voltage in conventional ST-FMR can be written
\begin{align}\label{Vmixlong}
    V^{\text{mix}}_{XX} = \frac{\Irf}{2}\Ramr \Re{m_x} \sin2\phi,
\end{align}
or
\begin{align}\label{Vmix2}\
\begin{split}
    V^{\text{mix}}_{XX} &= \frac{\Irf\Ramr}{2\alpha \omega^+} \sin(2\phi)\cos(\phi) \\
   &\quad\quad\quad\times \left( S(B) \tau^0_{\text{DL}}+A(B)\frac{\omega_2}{\omega}\tau^0_z \right)
\end{split}
\end{align}
where we have defined the symmetric Lorentzian $S(B)=\Delta^2/[(B-B_0)^2 + \Delta^2]$, the antisymmetric Lorentzian $A(B)=(B-B_0)\Delta/[(B-B_0)^2 + \Delta^2]$ and the half-width at half-maximum linewidth $\Delta = \alpha \omega / \gamma$.  Here $R_{\text{AMR}}$ includes contributions from both the anisotropic magnetoresistance in the magnet and the spin Hall magnetoresistance in the Pt layer, as these produce identical contributions to the ST-FMR signals for our sample geometry (see Supplementary Information \cite{Supplement}).

We can compute the transverse spin-torque mixing voltage within the same framework. We assume that the Hall resistance has the symmetry
$R_{XY} = R_{\text{PHE}}m_X m_Y + R_{\text{AHE}} m_Z$,
where $R_{\text{PHE}}$ is the scale of the planar Hall effect and $R_{\text{AHE}}$ is the scale of the anomalous Hall effect, in which case \cite{Hayashi2014}
\begin{align}\label{hayashi}
    V^{\text{mix}}_{XY}  &= \frac{\Irf}{2}\left( -\Rphe \cos 2\phi \, \Re{m_x} + \Rahe \Re{m_z}\right).
\end{align}
Using the results from Eq.\ (\ref{mymz}), 
\begin{align}\
\begin{split}
    V^{\text{mix}}_{XY} &= -\frac{\Irf\Rphe}{2\alpha \omega^+} \cos{(2\phi)} \cos(\phi) \\ 
   &\quad\quad\quad \times \left(S(B) \tau^0_{\text{DL}}+A(B)\frac{\omega_2}{\omega}\tau^0_z \right) \\
    &+ \frac{\Irf\Rahe}{2\alpha \omega^+}  \cos(\phi) \\
    &\quad\quad\quad \times \left(S(B) \tau^0_z-A(B)\frac{\omega_1}{\omega}\tau^0_{\text{DL}} \right).
\end{split}
\end{align}
\par
The artifact signals due to spin pumping and resonant heating can also contribute to both the longitudinal and transverse ST-FMR voltages \cite{Kondou2016,Okada2019,Kumar2019}. All of the artifacts we consider, SP/ISHE, LSSE/ISHE, and NE, produce resonant DC electric fields that are in-plane and perpendicular to the magnetization axis, and proportional to the square of the precession amplitude (with the precession amplitude $\propto \cos\phi$).  Because these signals depend only on the precession amplitude and not phase, they have symmetric lineshapes.  Taking the components in the longitudinal and transverse directions, the artifact voltages are therefore 
\begin{align}
\begin{split} \label{eq:Vsp}
	V_{\text{art}} = E^0_{\text{art}} S(B) \cos^2\phi 
		 \begin{dcases}
	     L\sin\phi & \text{longitudinal}\\
	     W\cos\phi & \text{transverse}
	 \end{dcases}
\end{split}
\end{align}
where $E^0_{\text{art}} = E^0_{\text{SP}} + E^0_{\text{LSSE}} + E^0_{\text{NE}}$ is the total electric field generated by all artifact signals. The artifact voltages for the longitudinal and transverse measurements differ only by geometric factors and angular symmetry: $L$ is the device length (parallel to the current flow) and $W$ is the transverse device width.
\par
The electric field due to the spin pumping/inverse spin Hall effect can be calculated by the method of ref.\ \cite{Tserkovnyak2002, Mosendz2010a} (see Supplementary Information \cite{Supplement}) 
\begin{align}\label{E0sp}
\begin{split}
	E^0_{\text{SP}} &= \frac{e \theta_\text{SH} \geff}{2\pi\sum_i \sigma_i t_i} \lsd \tanh\left(\frac{\tnm}{2\lsd}\right) \times \\ &\left[\frac{(\tau^0_\text{DL})^2 \omega_1+(\tau^0_z)^2 \omega_2}{\alpha^2  \left(\omega^+\right)^2}\right].
\end{split}
\end{align}
Here $\theta_\text{SH}$ is the spin Hall ratio in the HM (related to the damping-like spin torque efficiency by $\theta_\text{SH} = \xi_\text{DL}/T_\text{int}$, where $T_\text{int}$ is an interfacial spin transmission factor), $\geff$ is the real part of the effective spin mixing conductance, $\sigma_i$ ($t_i$) the charge conductivity (thickness) of layer $i$, and $\lsd$ the spin diffusion length of the HM. 
\par
If one assumes that the artifacts due to resonant heating by the current-induced torques are proportional to the energy absorbed by the magnetic layer during resonant excitation, the peak DC electric field due to LSSE/ISHE and NE can be calculated similarly \cite{Holanda2017, kikkawa2013} (see Supplementary Information \cite{Supplement})
\begin{align}\label{E0sse}
	E^0_{\text{LSSE}} + E^0_{\text{NE}} = C \frac{M_s \tfm \alpha \omega^+}{2 \gamma \sum_i \sigma_i t_i} \left[\frac{(\tau^0_\text{DL})^2 \omega_1+(\tau^0_z)^2 \omega_2}{\alpha^2  \left(\omega^+\right)^2}\right]. 
\end{align}
Here $C$ is a material-dependent prefactor. Due to the factor of $\tfm \alpha \omega^+$ in the numerator, the resonant heating contributions  scale differently than the SP/ISHE as a function of FM thickness, damping, and measurement frequency.
\par
Adding the rectification and artifact contributions [and using that $\cos^2\phi\sin\phi=(\sin2\phi\cos\phi)/2$ and $\cos^3\phi=(\cos\phi+\cos2\phi\cos\phi)/2$], the amplitudes of the symmetric and antisymmetric components of the total longitudinal and transverse ST-FMR signals have the angular dependence 
\begin{align}
\begin{split}\label{eq:SnAwithAngles}
        S_{XX}(\phi) &=  S_{XX}^\text{AMR/art}\sin 2\phi\cos\phi \\
		A_{XX}(\phi) &= A_{XX}^\text{AMR} \sin 2\phi\cos\phi\\
		S_{XY}(\phi) &= S_{XY}^\text{PHE/art}  \cos2\phi\cos\phi + S_{XY}^\text{AHE/art}\cos\phi\\
		A_{XY}(\phi) &= A_{XY}^\text{PHE}  \cos2\phi\cos\phi
		+ A_{XY}^\text{AHE} \cos\phi
\end{split}
\end{align}
with the amplitude coefficients
\begin{align}
    \begin{split}\label{AScoefs}
        S_{XX}^\text{AMR/art} &= \frac{\Irf}{2\alpha\omega^+} R_\text{AMR}\tau^0_\text{DL} - \frac{L}{2}  E^0_\text{art}  \\
        &\equiv S_{XX}^\text{AMR} + V_\text{art} \\
		A_{XX}^\text{AMR} &= \frac{\Irf}{2\alpha\omega^+} R_\text{AMR}\frac{\omega_2}{\omega}\tau^0_z\\
		S_{XY}^\text{PHE/art} &= -\frac{\Irf}{2\alpha\omega^+} \Rphe\tau^0_\text{DL} - \frac{W}{2} E^0_\text{art} \\
		A_{XY}^\text{PHE} &= -\frac{\Irf}{2\alpha\omega^+}\Rphe\frac{\omega_2}{\omega}\tau^0_z \\
		S_{XY}^\text{AHE/art} &=  \frac{\Irf}{2\alpha\omega^+}\Rahe\tau^0_z - \frac{W}{2} E^0_\text{art} \\
		A_{XY}^\text{AHE} &= -\frac{\Irf}{2\alpha\omega^+} \Rahe\frac{\omega_1}{\omega} \tau^0_\text{DL}.
    \end{split}
\end{align}
One can see that all of the $S_{XX}$ and $S_{XY}$ rectification signals are contaminated by artifact voltages. If one measures just $S_{XX}$ and $A_{XX}$ for in-plane magnetic fields (as in conventional ST-FMR) there is no way to distinguish $\tau^0_\text{DL}$ from the artifact contributions. However, $\tau^0_\text{DL}$ appears by itself, without any artifact contamination, in the coefficient $A_{XY}^\text{AHE}$. One way to achieve a measurement of $\tau^0_\text{DL}$, free of these artifacts, is therefore to directly use the expression for $A_{XY}^\text{AHE}$  in Eq.\ (\ref{AScoefs}) along with careful calibration of $\Irf$, $\alpha$, and $\Rahe$.  The out-of-plane torque $\tau^0_z$ can similarly be determined from $A_{XX}^\text{AMR}$ or $A_{XY}^\text{PHE}$.  Alternatively, the expressions in Eq.\ (\ref{AScoefs}) also allow $E^0_\text{art}$ and the torque efficiencies $\xi_\text{DL}$ and $\xi_\text{FL}$ to be measured without calibrating $\Irf$, $\alpha$, and the the magnetoresistance scales by taking appropriate ratios to cancel prefactors.  We can do so using measurements of either the set of parameters $\{ S_{XX}^\text{AMR/art}, A_{XX}^\text{AMR}, S_{XY}^\text{AHE/art}, A_{XY}^\text{AHE} \}$ or $\{ S_{XY}^\text{PHE/art}, A_{XY}^\text{PHE}, S_{XY}^\text{AHE/art}, A_{XY}^\text{AHE} \}$.  We do not expect that the equations involving $R_\text{AMR}$ and $\Rphe$ are physically independent because anisotropic magnetoresistance and the planar Hall effect originate from the same microscopic mechanism.  Therefore if the assumptions of our model are correct these two strategies for taking ratios to cancel prefactors must agree modulo experimental noise.  We will perform both calculations, and test their agreement as a consistency check.
\par
First, using that on resonance $\omega=\sqrt{\omega_1 \omega_2}$ we calculate the ratio $\eta \equiv (\tau^0_\text{DL}/\tau^0_z)\sqrt{\omega_1/\omega_2}$ employing the pair of parameters $S$ and $A$  associated with each of the AMR, PHE, and AHE:
% \begin{subnumcases}{\eta = \frac{-A^{\text{AHE}}_{XY}}{S^{\text{AHE/ISHE}} + W (E_\text{ISHE}/2)} = \\}
%     \frac{S^{\text{PHE/ISHE}}_{XY} + W (E_\text{ISHE}/2)}{A^{\text{PHE}}_{XY}}\label{PHE}\\
%     \frac{S^{\text{AMR/ISHE}}_{XX} + L  (E_\text{ISHE}/2)}{A^{\text{AMR}}_{XX}}\label{AMR}
% \end{subnumcases}
\[
\eta = \frac{-A^{\text{AHE}}_{XY}}{S^{\text{AHE/art}}_{XY} + W (E_\text{art}/2)} = \\
\]
\begin{subequations}
\begin{alignat}{2}[left=\empheqlbrace]
 & \frac{S^{\text{PHE/art}}_{XY} + W (E_\text{art}/2)}{A^{\text{PHE}}_{XY}}\label{PHE}\\
 & \frac{S^{\text{AMR/art}}_{XX} + L  (E_\text{art}/2)}{A^{\text{AMR}}_{XX}}\label{AMR}
\end{alignat}
\end{subequations}
Using the measured amplitude coefficients, one can solve separately for $E_\text{art}$ using either Eq.\ (\ref{PHE}) or (\ref{AMR}), and check consistency.

It still remains to determine $\tau^0_\text{DL}$ and to separate the two contributions to $\tau^0_z = \tau^0_\text{FL}+\tau^0_\text{Oe}$.
We choose to do this using a method from ref.\ \cite{Pai2015}, in a way that determines both the of the spin-torque efficiencies $\xi_\text{DL}$ and $\xi_\text{FL}$ at the same time without requiring a separate calibration of $I_\text{RF}$.  We perform measurements for a series of samples with different thicknesses of the ferromagnetic layer and determine $\eta =(\tau^0_\text{DL}/\tau^0_z)\sqrt{\omega_1/\omega_2}$ for each sample from any of the expressions in Eqs.\ (\ref{PHE},\ref{AMR}), after solving for $E_\text{art}$.  We then define
\begin{align}\label{xifmr1}
		\xifmr &\equiv \eta\frac{e\mu_0M_s\tnm\tfm}{\hbar} \sqrt{1 + \frac{\mu_0\meff}{B_0}}
\end{align}
so that using Equations (\ref{taudl}) \& (\ref{taufl}), and that by Ampere's Law $\tau^0_\text{Oe} = \gamma \mu_0 J_e t_\text{HM}/2$ one has
\begin{align}\label{xifmr2}
        \frac{1}{\xifmr} &= \frac{1}{\xidl}\left( 1+ \frac{\hbar}{e} \frac{\xifl}{\mu_0 M_s \tfm \tnm }\right).
\end{align}
Performing a linear fit of $1/\xifmr$ vs.\ $1/t_\text{FM}$ then can be used to determine $1/\xi_\text{DL}$ (from the intercept) and $\xi_\text{FL}$ (from the slope).

\begin{figure*}
\begin{center}
\begin{tabular}{ccc}
\subfloat[]{\includegraphics[width = 0.33\linewidth]{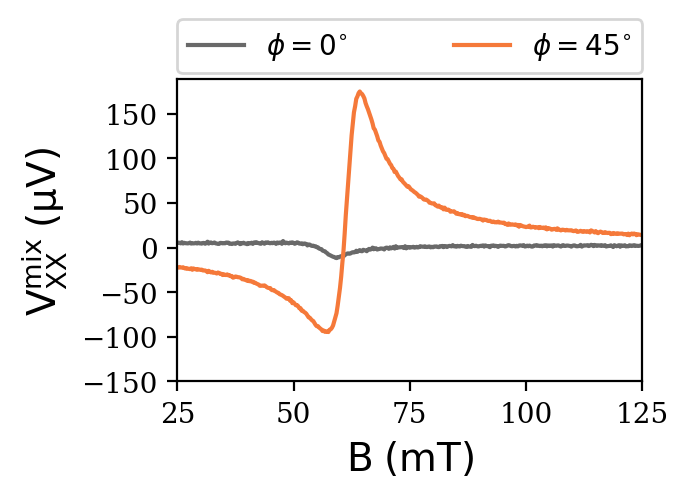}}&
\subfloat[]{\includegraphics[width = 0.33\linewidth]{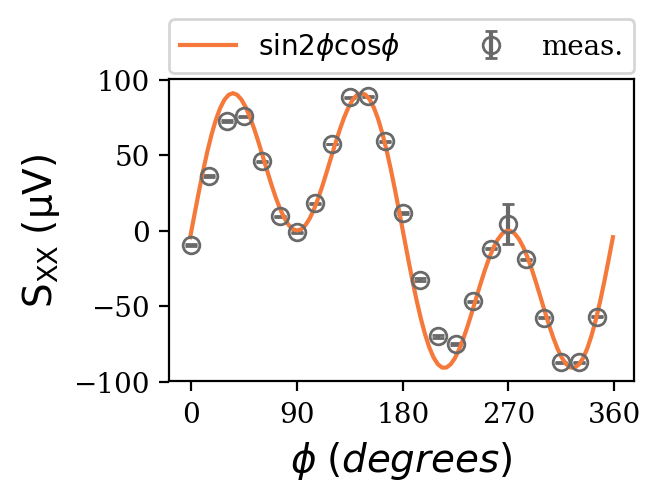}}&
\subfloat[]{\includegraphics[width = 0.33\linewidth]{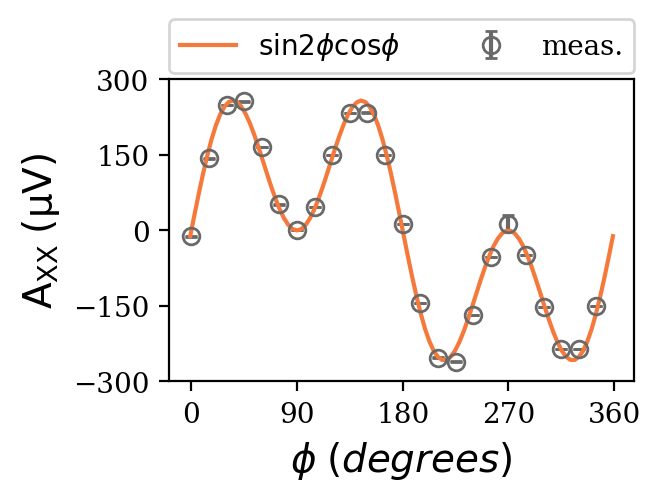}}\\
\subfloat[]{\includegraphics[width = 0.33\linewidth]{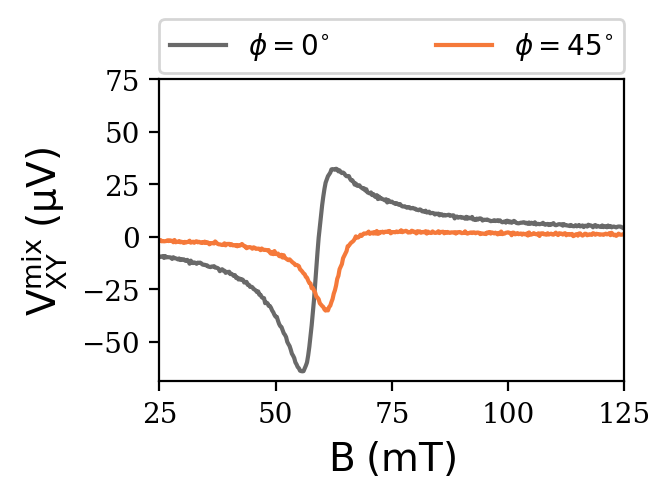}} &
\subfloat[]{\includegraphics[width = 0.33\linewidth]{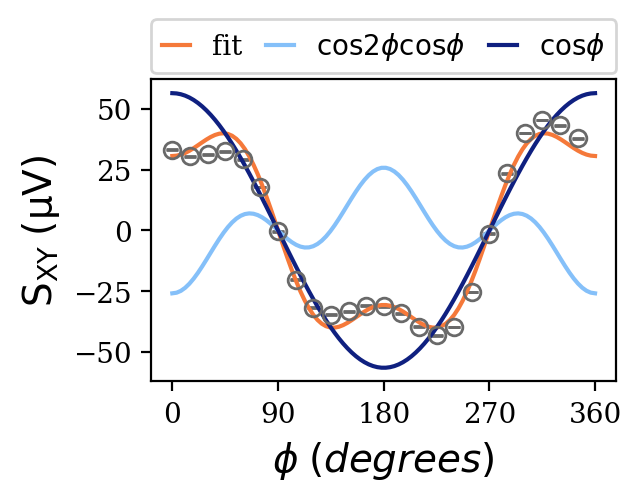}} &
\subfloat[]{\includegraphics[width = 0.33\linewidth]{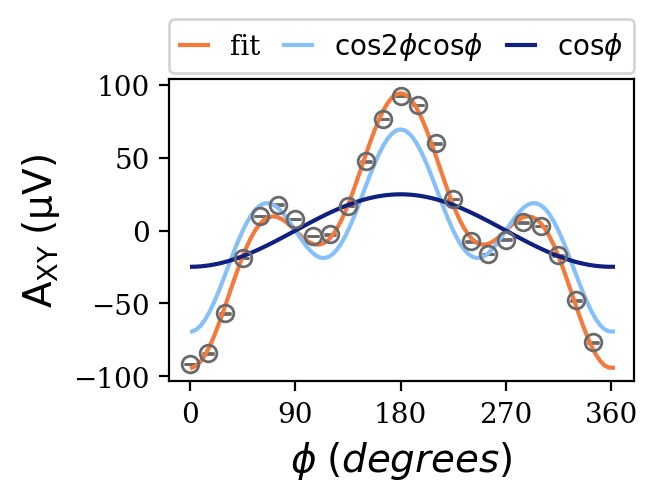}}
\end{tabular}
\caption{ST-FMR measurements of a Pt(6 nm)/CoFeB(6 nm) sample for a measurement frequency $f= 8$ GHz. \textbf{(a)} Longitudinal resonant signals for field sweeps with two different field angles. \textbf{(b) \& (c)} Symmetric ($S_{XX}$) and antisymmetric ($A_{XX}$) Lorentzian fit components for the longitudinal resonant signal as a function of the external field angle. \textbf{(d)} Transverse resonant signals for field sweeps with two different field angles. \textbf{(e) \& (f)} Symmetric ($S_{XY}$) and antisymmetric ($A_{XY}$) Lorentzian fit components for the transverse resonant signal as a function of the external field angle. The orange fit line in (b) \& (c) is a fit to $\sin2\phi\cos\phi$ (AMR); the light and dark blue fit lines in (e,f) are fits to $\cos2\phi\cos\phi$ (PHE) and $\cos\phi$ (AHE), respectively, and their sum (orange) fits the data.}
\label{fig:2}
\end{center}
\end{figure*}

%Another class of artifacts that can contribute to low-frequency, non-resonant measurements of spin-orbit torques are thermoelectric effects arising from current-induced heating \cite{Avci2014}.  However, since current-induced heating will be approximately equal for applied magnetic fields on and off resonance, the main consequence of thermoelectric effects for the resonant ST-FMR technique will be a shifted background, rather than a change in the resonant peak amplitude or shape.  Any small difference in heating due to microwave absorption at resonance will contribute a signal with the same angular dependence as the ISHE signal, so that we will not consider it separately in our analysis.

\section{Measurements}

We used DC-magnetron sputtering to grow multiayers with the structure substrate/Ta(1)/Pt(6)/ferromagnet($t_{\text{FM}}$)/Al(1) (where numbers in parentheses are thicknesses in nm), using three different ferromagnets (FMs): Co$_{40}$Fe$_{40}$B$_{20}$ (CoFeB), permalloy (Ni$_{81}$Fe$_{19}$ = Py) and Co$_{90}$Fe$_{10}$ (CoFe).  Each of the three FMs is expected to have different AMR, PHE, and AHE values, and therefore different strengths of rectified spin-torque signals relative to the artifacts.  In particular, CoFeB has weak planar magnetoresistances (AMR and PHE), and has been argued previously to exhibit a significant contribution from SP/ISHE in ST-FMR \cite{Kondou2016,Okada2019}. The CoFeB devices were grown with $t_{\text{FM}} = \{2,3,4,6,8,10\}$ in separate depositions.  The Py and Co$_{90}$Fe$_{10}$ devices were grown with single relatively-large thicknesses to give measurable artifact signals: $t_\text{Py}$= 8 nm and $t_\text{CoFe}$= 6 nm.  All devices were grown on high-resistivity ($>2\times10^4 \,\Omega\text{-cm}$), thermally-oxidized silicon wafers to prevent RF current leakage or capacitive coupling. The Ta was used as a seed layer and has negligible contribution to the SOTs we measure due to the low conductivity of Ta relative to Pt ($\rho_{\text{Pt}}$ = 20.4 $\mu \Omega$cm, $\rho_{\text{CoFeB}}$ = 110 $\mu \Omega$cm). The Al cap layer protects the layers below it, and is oxidized upon exposure to atmosphere. 
\par
The as-deposited samples were patterned using photolithography and Ar ion-milling to define rectangular bars ranging in size from 20 $\times$ 40 $\mu$m to 40 $\times$ 80 $\mu$m with various aspect ratios. The transverse leads and contact pads were then made using a second photolithography step, deposited by sputtering Ti(3 nm)/Pt(75 nm) and formed by lift-off so that the side channels extended a few microns on top of the main bar (see Fig.\ \ref{fig:1}). We were careful that the magnetic layer did not extend beyond the defined rectangle into the transverse leads.  In early devices, we etched full Hall-bar shapes within the first layer of lithography so that the transverse leads included some of the same magnetic layer as the main channel.  For those early devices, we found that the resulting analyses of spin-orbit torques produced anomalous results, varying with the dimensions of the leads and the contact separation.  This could possibly be due to spatial non-uniformities in the magnetic orientation  and precession, as was speculated in \cite{bose2017}.  Ultimately,  the magnetic bilayer was left to be simply rectangular to promote uniform precession modes, and this removed the anomalous geometry dependence.
\par
For the ST-FMR measurements, we connected the devices to an amplitude-modulated (``AM" with $f_{\text{AM}}\approx$ 1700 Hz) microwave source through the AC port of a bias tee and to a lock-in amplifier through the DC port, which detected the longitudinal signal. Another lock-in amplifier measured the DC voltage across the Hall leads of the device. Both lock-in amplifiers referenced the same AM signal, and we collected ST-FMR data in both the longitudinal and transverse directions simultaneously.  An in-plane applied magnetic field was applied at varying angles $\phi$ using a projected-field magnet.  We used fixed microwave frequencies in the range 7-12 GHz, applied 20 dBm of microwave power, and all measurements were performed at room temperature. In Figs.\ \ref{fig:2}(a) and \ref{fig:2}(d) we show examples of the detected resonant signals from the parallel ($XX$) and transverse ($XY$) lock-ins for the Pt(6)/CoFeB(6) sample.
\par
 Both the longitudinal and transverse resonances are well-fit to a sum of symmetric and antisymmetric Lorentzian peaks, with varying relative weights.  For each sample we performed field-swept measurements at a variety of angles $\phi$, extracting the symmetric and antisymmetric components of the resonances for both the longitudinal and transverse signals. The results for a Pt(6)/CoFeB(6) sample are shown in Fig.\ \ref{fig:2}(b,c,e,f), along with fits to Eq.\ (\ref{eq:SnAwithAngles}).  Analogous results for Pt(6)/Py(8) and Pt(6)/CoFe(6) samples are shown in the Supplementary Information \cite{Supplement}.  
 \par
 We find excellent agreement with the expected angular dependences for $S_\text{XX}$, $A_\text{XX}$, and $A_\text{XY}$.  For $S_\text{XY}$ the dominant contributions to the angular dependence are, as expected the $\cos2\phi\cos\phi$ and $\cos\phi$ terms, but in addition, we detect a small component approximately proportional to $\sin2\phi$.  This additional contribution is less than 10\% of the larger terms in $S_\text{XY}$ for all thicknesses of CoFeB, small enough that it is not included in the fit shown in Fig.\ \ref{fig:2}(e). It is more significant in the CoFe and Py samples that we measured, though still smaller than the $\cos2\phi\cos\phi$ and $\cos\phi$ amplitudes in $S_\text{XY}$ (see Supplementary Information \cite{Supplement}). A $\sin2\phi$ contribution can only arise from a breaking of mirror symmetry relative to the sample's $\hat{Y}$-$\hat{Z}$ plane (see Supplementary Information \cite{Supplement}).  This symmetry is broken in our samples by the different contact geometries on the two ends of the sample wire (see Fig. 1(a)).  The form of the $\sin2\phi$ signal can be explained as due resonant heating that produces an in-plane thermal gradient in the longitudinal direction of the sample (due {\it e.g.\ }to differences in heat sinking at the two ends) that is transduced to a tranverse voltage with the symmetry of the planar Hall effect ($\propto m_X m_Y$).  We have checked that the signal is not due to a sample tilt or to a non-resonant DC current that might arise from rectification of the applied microwave signal at the sample contacts.  All of the other Fourier components that are the main subject of our analysis maintain the $\hat{Y}$-$\hat{Z}$-plane mirror symmetry, and so they cannot be altered at first order by a process that breaks this symmetry.  Being a separate Fourier component, the $\sin2\phi$ contribution also does not affect the fits to Eq.\ (\ref{eq:SnAwithAngles}) to determine the six amplitude coefficients
         $S_{XX}^\text{AMR/art}$, 
		$A_{XX}^\text{AMR}$,
		$S_{XY}^\text{PHE/art}$, 
		$A_{XY}^\text{PHE}$, 
		$S_{XY}^\text{AHE/art}$, and
		$A_{XY}^\text{AHE}$.
 Using these coefficients, we  calculate $E_\text{art}$ by solving Eqs.\ (\ref{PHE}) or (\ref{AMR}). There is a potential ambiguity in which roots of Eqs.\ (\ref{PHE}) and (\ref{AMR}) to select when applying the quadratic formula. In our measurements, one root would give unphysical results, e.g.\ a sign change of $\xidl$.  An important check of our method (and a check that the $\sin2\phi$ term in $S_{XY}$ does not contaminate the analysis) is that these two independent methods for determining $E^0_\text{art}$ (Eqs.\ (\ref{PHE}) and (\ref{AMR})) give consistent results.  We show below that this is indeed the case.

\begin{center}
\begin{figure}[h]
\begin{tabular}{c}
\subfloat[]{\includegraphics[width = \linewidth]{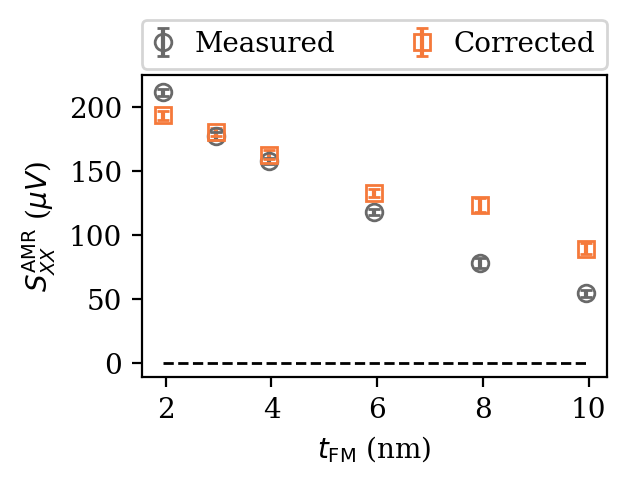}}\\
\subfloat[]{\includegraphics[width = \linewidth]{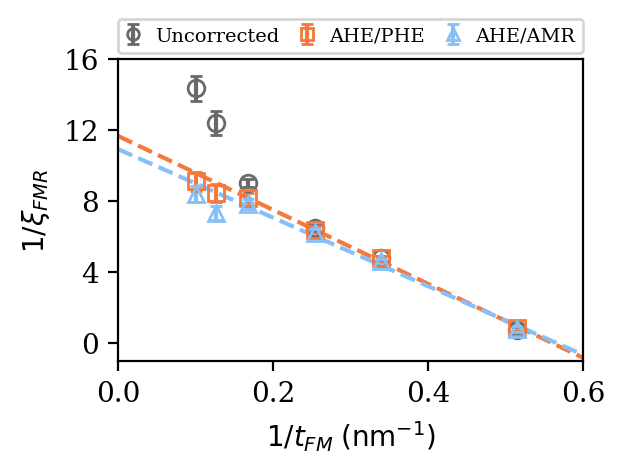}}
\end{tabular}
\caption{\textbf{(a)} The uncorrected measured value of $S_{XX}^\text{AMR}$ vs. $\tfm$, together with the value corrected by removing the artifact voltage. 
%The shaded area is calculated from Eq. (\ref{AScoefs}) using measured values with no adjustable parameters and indicating a $\pm2\sigma$ region taking into account the error bars of $\xidl$ and $I_\text{RF}$. 
\textbf{(b)} The inverse $\xifmr$ vs. inverse $\tfm$. The y-intercept of the line is $1/\xidl$ and the slope is proportional to $\xifl$ as in Eq. (\ref{xifmr2}). The two fit lines are color-matched fits to the data points from the AHE/PHE and AHE/AMR corrections.}
\label{fig:3}
\end{figure}
\end{center}

\begin{center}
\begin{figure}[h]
\includegraphics[width = \linewidth]{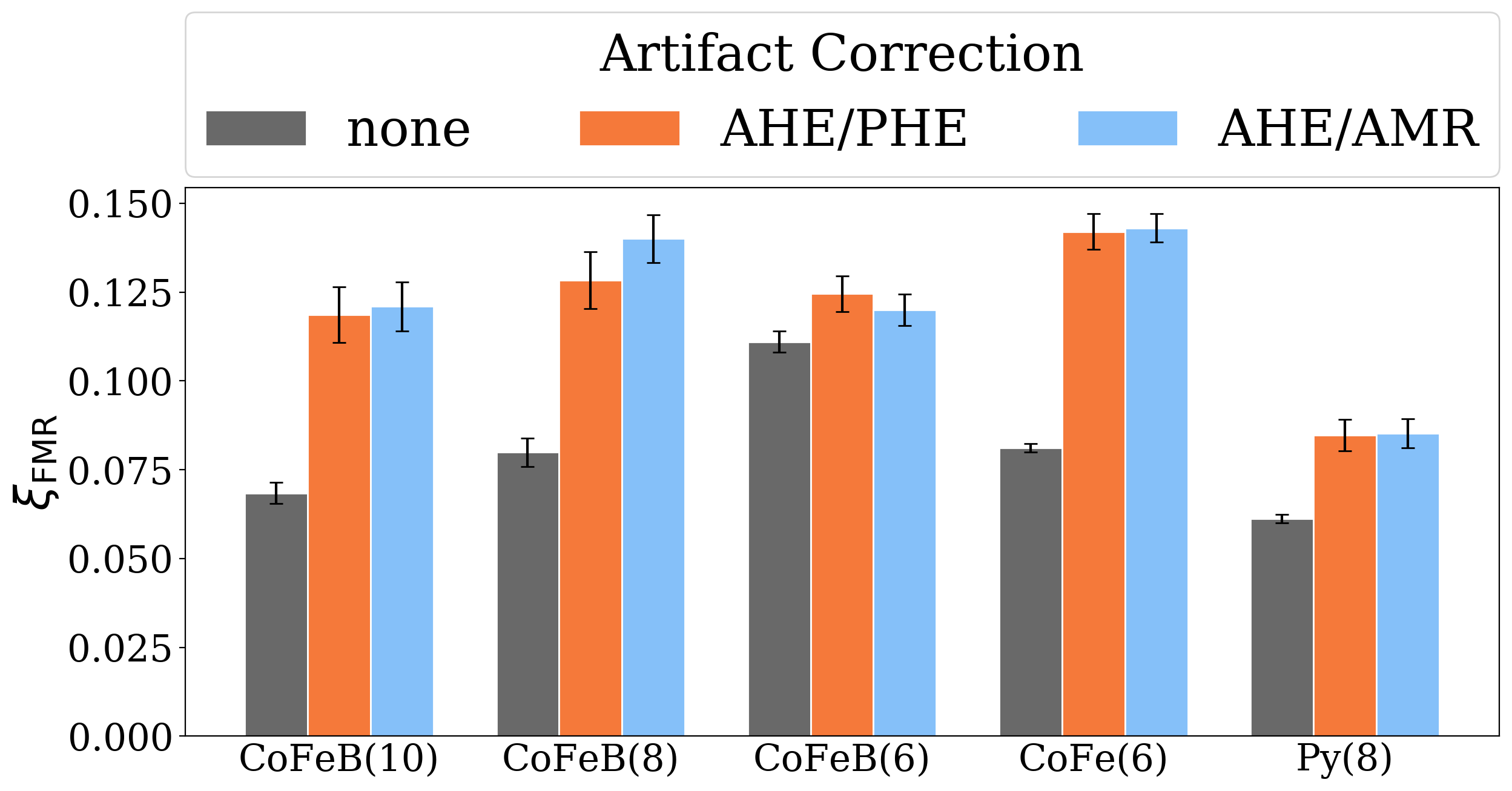}\\
\caption{$\xifmr$ for various device stacks. The gray (left) bars show  values without correction for the artifacts, and the orange and blue (center, right) bars show values corrected using the determination of the artifact voltages using Eqs.\ (\ref{PHE}) and  (\ref{AMR}), respectively.}
\label{fig:4}
\end{figure}
\end{center}

Figure \ref{fig:3}(a) shows the total amplitude of the longitudinal symmetric ST-FMR component (labeled as ``Measured''), and the corrected value $S^{\text{AMR}}_\text{XX}$ from which  $V_{\text{art}}$ has been subtracted.    For CoFeB layer thicknesses 6 nm and below, the magnitude of $V_\text{art}$ is much less than the magnitude of $S^{\text{AMR}}_\text{XX}$, so that the artifacts have little effect on ST-FMR measurements of the spin-orbit torques.  However, with increasing CoFeB thickness the magnitude of $S^{\text{AMR}}_\text{XX}$ decreases and $V_\text{art}$ grows, so we find experimentally that for the CoFeB layers thicker than 6 nm the artifact voltage becomes a significant fraction of the total signal.  In this regime, $V_\text{art}$ and $S^{\text{AMR}}_\text{XX}$ contribute to $S_{\text{XX}}(\phi)$ with opposite signs \cite{schreier2014}, with the consequence that if the artifact contributions are neglected in the conventional ST-FMR analysis, the result is an underestimate of the strength of $\tau^0_\text{DL}$. In this respect our results conflict with some conclusions \cite{Kondou2016,Okada2019} that neglecting the SP/ISHE contribution produces an overestimate of $\tau^0_\text{DL}$. 
%The shaded area in Fig.\ \ref{fig:3}(a) shows the expected thickness dependence of the rectified longitudinal symmetric signal, $S^{\text{AMR}}_\text{XX}$, in the absence of the artifacts, according to the formula
%\begin{align}\label{SxxFit}
%S_{XX}^\text{AMR} &= \frac{\Irf R_\text{AMR}}{2\alpha\omega^+} %\tau^0_\text{DL}
%\end{align}
%with $\pm2\sigma$ errorbars on both $\xidl$ and $I_\text{RF}$ defining the shaded region. Here, we use the value of $\tau^0_\text{DL}$ determined below, so that the comparison can be made with no free parameters.  We find reasonable agreement with the expected thickness dependence of the rectification signal.
\par
Analysis of the dependence of $1/\xi_\text{FMR}$ as a function of $1/t_\text{FM}$ allows a determination of the underlying spin-torque efficiencies $\xi_\text{DL}$ and $\xi_\text{FL}$ using Eq.\ (\ref{xifmr2}).  The results for the CoFeB series of samples is shown in Fig.\ 3(b).  If one does not correct for the contribution of the artifacts, the calculated values of  $1/\xi_\text{FMR}$ depart upward from the expected linear dependence for $t_\text{FM} \gtrsim 6$ nm.  Similar results have been reported previously in \cite{Pai2015} where the non-linearity was speculated to be from SP/ISHE, and the spin-torque efficiencies were determined by fitting only to the thinner FM stacks.  After we correct for the artifact contribution, we find good agreement with the expected linear dependence over the full thickness range.  From the linear fit, we determine  $\xi_\text{DL}=$ 0.090(6) and $\xi_\text{FL}=$ -0.020(2). 
\par
% As a check, we can determine $\xi_\text{DL}$ from the $A^\text{AHE}_{XY}$ using Eqs.\ (\ref{AScoefs}) \& (\ref{taudl}), along with a separate calibration of $I_\text{RF}$. Calibrating the microwave current using a vector network analyzer and measuring $R_{\text{AHE}}$, we find that $\xidl = 0.075(8)$, in reasonable agreement with the results of the fit in Fig.\ \ref{fig:5}. 
\par
For the Pt(6 nm)/Py(8 nm) and Pt(6 nm)/CoFe(6 nm) samples we find the same configuration of signs as for the thicker Pt/CoFeB samples: $V_\text{art}$ partially cancels $S^{\text{AMR}}_\text{XX}$ so that the true mixing signal is larger than the measured amplitude of $S_{\text{XX}}(\phi)$. The results of the calculation of $\xi_\text{FMR}$ according to Eq.\ (\ref{xifmr1}) are shown in Fig.\ \ref{fig:4} for five selected samples, both without and with the correction for artifacts.  In determining $\xi_\text{FMR}$ we use values for $M_s$ determined by room temperature vibrating sample magnetometry (VSM) and values for $\mu_0 M_\text{eff}$ determined by fits of the ST-FMR resonant fields as a function of frequency. These values are: for CoFeB $M_s= 9.8 \times 10^5$ A/m, $\mu_0 M_\text{eff}= $ 0.6 -- 1.4 T (depending on thickness); for Py $M_s=  7.5 \times 10^5$ A/m, $\mu_0 M_\text{eff}= 1.01$ T; and for CoFe $M_s= 9.1\times 10^5$ A/m, $\mu_0 M_\text{eff}= 1.66$ T. If a magnetic dead layer was observed in VSM, the dead layer thickness was subtracted from $\tfm$. In all cases shown in Fig.\ 4, we find that correcting for the artifact contribution increases our estimates for the values of $\xi_\text{FMR}$. The value of $\xi_\text{FMR}$ is smaller for the Pt/Py sample than for Pt/CoFeB or Pt/CoFe primarily because $\xi_\text{FL}$ is both small and has a positive sign for Pt/Py \cite{fan2013,nan2015}.
\par
The dependence of the artifact voltage, $V_\text{art}$, on the ferromagnetic layer thickness is shown in Fig.\ 5 for the longitudinal ST-FMR component of the Pt/CoFeB series of samples.
The data are compared to an estimate of the SP/ISHE contribution from Eq.\ (\ref{E0sp}), using the parameters (appropriate for the resistivity of our Pt layers, $\rho_{\text{Pt}}=20.4$ $\mu \Omega$cm): $\theta_\text{SH} = 0.32$ \cite{Zhu2019geff,Pai2015}, $\geff = 8.26\times 10^{18}$ m$^{-2}$ \cite{Zhu2019geff}, and $\lsd = 3.7$ nm \cite{Nguyen2016}. The other quantities in Eq.\ (\ref{E0sp}) were measured for our samples, including the variation as a function of CoFeB thickness. The comparison therefore includes no adjustable fitting parameters, but given that there is considerable disagreement in the literature about the values of the parameters $\theta_\text{SH}$, $\geff$, and $\lsd$, one should still be careful about drawing quantitative conclusions.  The comparison indicates to us that for the samples with $t_{\text{FM}} \geq$ 3 nm  the SP/ISHE theory predicts the correct sign and can roughly capture the overall magnitude and thickness-dependence of the measured artifact signal. However, the measured artifact voltage for $t_\text{FM}$= 2 nm has the opposite sign, inconsistent with the SP/ISHE.  We are confident that the measured sign change is real, because we have measured and performed the analysis on five Pt(6 nm)/CoFeB(2 nm) devices with varied geometries, with consistent results. 

\begin{center}
\begin{figure}[t]
\begin{tabular}{c}
\includegraphics[width = \linewidth]{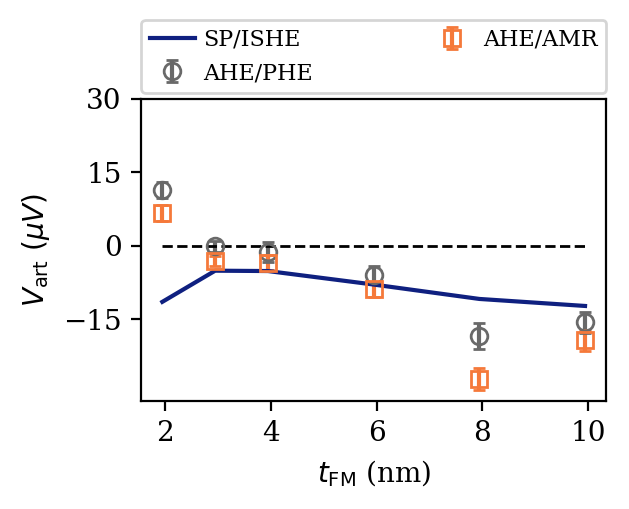}\\
\end{tabular}
\caption{The artifact voltage as a function of the FM thickness in Pt(6 nm)/CoFeB samples. The two types of data points reflect the two correction equations ((\ref{PHE}) and (\ref{AMR})). The line is the estimated SP/ISHE contribution, determined using the parameters described in the text, with no adjustable parameters.}
\label{fig:5}
\end{figure}
\end{center}
\par
Given that the SP/ISHE cannot explain the sign change in the artifact voltage for our $t_\text{FM}$= 2 nm samples, we suggest that resonant heating effects might be comparable to the SP/ISHE in our Pt(6 nm)/CoFeB samples, with sufficient strength to reverse the overall sign of the artifact voltage for our thinnest samples.  This suggestion differs from previous studies on Pt/YIG samples, for which frequency-dependent measurements demonstrated that SP/ISHE signals dominate over resonant heating artifacts \cite{Iguchi2017,Iguchi2012}. However, the relative strength of the heating effects and SP/ISHE should scale proportional to the damping $\alpha$ (compare Eqs.\ (\ref{E0sp}) and (\ref{E0sse})), so that the heating effects should be more significant in higher-damping ferromagnetic metals compared to lower-damping YIG.  We calculate that the resonant heating due to the excitation of magnetic precession for our 2 nm samples is $\sim 2.5 \times 10^4$ Wm$^{-2}$ (Supplementary Information \cite{Supplement}), only about a factor of 5 less than the Ohmic heating per unit area in the CoFeB, $\sim 1.2 \times 10^5$ Wm$^{-2}$. 
%Given that the resonant heating deposits energy only in the magnetic layer, rather than in both metallic layers, and in fact mainly into the spin degrees of freedom in the magnetic layer, 
We suggest that this is sufficient to measurably alter the thermal gradients within the sample at resonance and induce resonant signals from the LSSE and/or Nernst effects. Due to an increase in the damping coefficient $\alpha$ with decreasing magnetic thickness, the ratio of the resonant heating to Ohmic heating is significantly greater for the 2 nm CoFeB samples than for the thicker magnetic layers (see Supplmentary Information \cite{Supplement}).

\par
As noted in the introduction, past experiments have shown a discrepancy between measurements of $\xidl$ using low frequency second harmonic Hall and ST-FMR techniques. To see if our correction for the artifact voltages in ST-FMR alleviates the discrepancy between the two techniques, we carried out low frequency second harmonic Hall measurements on the same Pt/CoFeB bilayers \cite{Supplement}.  We found that the low frequency second harmonic measurements of $\xidl$ were still approximately 60\% larger than what we measured by ST-FMR, even after correcting ST-FMR for spin pumping and resonant heating.  This persisting quantitative difference suggests that the assumptions used in analyzing one or both of these experiments are missing an important bit of physics.  Our analysis indicates that this missing physics is not simply the neglect of spin pumping or a simple heating-induced voltage in the ST-FMR results, and therefore more work must be done to understand the source of the disagreement.

\section{Conclusion}
In conclusion, we have demonstrated that the rectification signal used to measure the strength of spin-orbit torques in spin-torque ferromagnetic resonance (ST-FMR) can be separated from artifact voltages that may arise due to spin pumping and resonant heating by performing ST-FMR in the transverse (Hall) configuration as well as the usual longitudinal configuration.  For Pt(6 nm)/CoFeB($t_{\text{FM}}$) samples, the artifact voltages are small compared to the rectification signal for $t_{\text{FM}}<$ 6 nm, but they can become a significant part of the measured signal for thicker magnetic layers.  The sign and overall magnitude of the measured artifact voltage for these thicker layers are consistent with expectations for the SP/ISHE effect signal.  However, the sign of the artifact voltage is reversed for our thinnest magnetic layers, with $t_{\text{FM}}$ = 2 nm.  This sign reversal cannot be explained by the SP/ISHE, so we suggest that it may be caused by a resonant heating effect.    

\section{Acknowledgements}
This research was supported in part by task 2776.047 in ASCENT, one of six centers in JUMP, a Semiconductor Research Corporation program sponsored by DARPA, and in part by the National Science Foundation (DMR-1708499). The devices were fabricated using the shared facilities of the Cornell NanoScale Facility, a member of the National Nanotechnology Coordinated Infrastructure (NNCI), and the Cornell Center for Materials Research, both of which are supported by the NSF (NNCI-1542081 and DMR-1719875).

\newpage

%\printbibliography
\bibliographystyle{apsrev4-1}
\bibliography{bibl}\clearpage

\begin{thebibliography}{51}%
\makeatletter
\providecommand \@ifxundefined [1]{%
 \@ifx{#1\undefined}
}%
\providecommand \@ifnum [1]{%
 \ifnum #1\expandafter \@firstoftwo
 \else \expandafter \@secondoftwo
 \fi
}%
\providecommand \@ifx [1]{%
 \ifx #1\expandafter \@firstoftwo
 \else \expandafter \@secondoftwo
 \fi
}%
\providecommand \natexlab [1]{#1}%
\providecommand \enquote  [1]{``#1''}%
\providecommand \bibnamefont  [1]{#1}%
\providecommand \bibfnamefont [1]{#1}%
\providecommand \citenamefont [1]{#1}%
\providecommand \href@noop [0]{\@secondoftwo}%
\providecommand \href [0]{\begingroup \@sanitize@url \@href}%
\providecommand \@href[1]{\@@startlink{#1}\@@href}%
\providecommand \@@href[1]{\endgroup#1\@@endlink}%
\providecommand \@sanitize@url [0]{\catcode `\\12\catcode `\$12\catcode
  `\&12\catcode `\#12\catcode `\^12\catcode `\_12\catcode `\%12\relax}%
\providecommand \@@startlink[1]{}%
\providecommand \@@endlink[0]{}%
\providecommand \url  [0]{\begingroup\@sanitize@url \@url }%
\providecommand \@url [1]{\endgroup\@href {#1}{\urlprefix }}%
\providecommand \urlprefix  [0]{URL }%
\providecommand \Eprint [0]{\href }%
\providecommand \doibase [0]{http://dx.doi.org/}%
\providecommand \selectlanguage [0]{\@gobble}%
\providecommand \bibinfo  [0]{\@secondoftwo}%
\providecommand \bibfield  [0]{\@secondoftwo}%
\providecommand \translation [1]{[#1]}%
\providecommand \BibitemOpen [0]{}%
\providecommand \bibitemStop [0]{}%
\providecommand \bibitemNoStop [0]{.\EOS\space}%
\providecommand \EOS [0]{\spacefactor3000\relax}%
\providecommand \BibitemShut  [1]{\csname bibitem#1\endcsname}%
\let\auto@bib@innerbib\@empty
%</preamble>
\bibitem [{\citenamefont {Brataas}\ \emph {et~al.}(2012)\citenamefont
  {Brataas}, \citenamefont {D~Kent},\ and\ \citenamefont {Ohno}}]{Brataas2012}%
  \BibitemOpen
  \bibfield  {author} {\bibinfo {author} {\bibfnamefont {A.}~\bibnamefont
  {Brataas}}, \bibinfo {author} {\bibfnamefont {A.}~\bibnamefont {D~Kent}}, \
  and\ \bibinfo {author} {\bibfnamefont {H.}~\bibnamefont {Ohno}},\ }\href
  {\doibase 10.1038/nmat3311} {\bibfield  {journal} {\bibinfo  {journal}
  {Nature Materials}\ }\textbf {\bibinfo {volume} {11}},\ \bibinfo {pages}
  {372} (\bibinfo {year} {2012})}\BibitemShut {NoStop}%
\bibitem [{\citenamefont {Wang}\ \emph {et~al.}(2013)\citenamefont {Wang},
  \citenamefont {Alzate},\ and\ \citenamefont {Amiri}}]{Wang2013}%
  \BibitemOpen
  \bibfield  {author} {\bibinfo {author} {\bibfnamefont {K.~L.}\ \bibnamefont
  {Wang}}, \bibinfo {author} {\bibfnamefont {J.~G.}\ \bibnamefont {Alzate}}, \
  and\ \bibinfo {author} {\bibfnamefont {P.~K.}\ \bibnamefont {Amiri}},\ }\href
  {\doibase 10.1088/0022-3727/46/7/074003} {\bibfield  {journal} {\bibinfo
  {journal} {Journal of Physics D: Applied Physics}\ }\textbf {\bibinfo
  {volume} {46}},\ \bibinfo {pages} {074003} (\bibinfo {year}
  {2013})}\BibitemShut {NoStop}%
\bibitem [{\citenamefont {{Oboril}}\ \emph {et~al.}(2015)\citenamefont
  {{Oboril}}, \citenamefont {{Bishnoi}}, \citenamefont {{Ebrahimi}},\ and\
  \citenamefont {{Tahoori}}}]{Oboril2015}%
  \BibitemOpen
  \bibfield  {author} {\bibinfo {author} {\bibfnamefont {F.}~\bibnamefont
  {{Oboril}}}, \bibinfo {author} {\bibfnamefont {R.}~\bibnamefont {{Bishnoi}}},
  \bibinfo {author} {\bibfnamefont {M.}~\bibnamefont {{Ebrahimi}}}, \ and\
  \bibinfo {author} {\bibfnamefont {M.~B.}\ \bibnamefont {{Tahoori}}},\ }\href
  {\doibase 10.1109/TCAD.2015.2391254} {\bibfield  {journal} {\bibinfo
  {journal} {IEEE Transactions on Computer-Aided Design of Integrated Circuits
  and Systems}\ }\textbf {\bibinfo {volume} {34}},\ \bibinfo {pages} {367}
  (\bibinfo {year} {2015})}\BibitemShut {NoStop}%
\bibitem [{\citenamefont {Liu}\ \emph {et~al.}(2011)\citenamefont {Liu},
  \citenamefont {Moriyama}, \citenamefont {Ralph},\ and\ \citenamefont
  {Buhrman}}]{Liu2011}%
  \BibitemOpen
  \bibfield  {author} {\bibinfo {author} {\bibfnamefont {L.}~\bibnamefont
  {Liu}}, \bibinfo {author} {\bibfnamefont {T.}~\bibnamefont {Moriyama}},
  \bibinfo {author} {\bibfnamefont {D.~C.}\ \bibnamefont {Ralph}}, \ and\
  \bibinfo {author} {\bibfnamefont {R.~A.}\ \bibnamefont {Buhrman}},\ }\href
  {\doibase 10.1103/PhysRevLett.106.036601} {\bibfield  {journal} {\bibinfo
  {journal} {Phys. Rev. Lett.}\ }\textbf {\bibinfo {volume} {106}},\ \bibinfo
  {pages} {036601} (\bibinfo {year} {2011})}\BibitemShut {NoStop}%
\bibitem [{\citenamefont {Liu}\ \emph {et~al.}(2012)\citenamefont {Liu},
  \citenamefont {Pai}, \citenamefont {Li}, \citenamefont {Tseng}, \citenamefont
  {Ralph},\ and\ \citenamefont {Buhrman}}]{Liu2012}%
  \BibitemOpen
  \bibfield  {author} {\bibinfo {author} {\bibfnamefont {L.}~\bibnamefont
  {Liu}}, \bibinfo {author} {\bibfnamefont {C.-F.}\ \bibnamefont {Pai}},
  \bibinfo {author} {\bibfnamefont {Y.}~\bibnamefont {Li}}, \bibinfo {author}
  {\bibfnamefont {H.~W.}\ \bibnamefont {Tseng}}, \bibinfo {author}
  {\bibfnamefont {D.~C.}\ \bibnamefont {Ralph}}, \ and\ \bibinfo {author}
  {\bibfnamefont {R.~A.}\ \bibnamefont {Buhrman}},\ }\href
  {https://science.sciencemag.org/content/336/6081/555} {\bibfield  {journal}
  {\bibinfo  {journal} {Science}\ }\textbf {\bibinfo {volume} {336}},\ \bibinfo
  {pages} {555} (\bibinfo {year} {2012})}\BibitemShut {NoStop}%
\bibitem [{\citenamefont {Mellnik}\ \emph {et~al.}(2014)\citenamefont
  {Mellnik}, \citenamefont {Lee}, \citenamefont {Richardella}, \citenamefont
  {Grab}, \citenamefont {Mintun}, \citenamefont {Fischer}, \citenamefont
  {Vaezi}, \citenamefont {Manchon}, \citenamefont {Kim}, \citenamefont
  {Samarth},\ and\ \citenamefont {Ralph}}]{Mellnik2014}%
  \BibitemOpen
  \bibfield  {author} {\bibinfo {author} {\bibfnamefont {A.~R.}\ \bibnamefont
  {Mellnik}}, \bibinfo {author} {\bibfnamefont {J.~S.}\ \bibnamefont {Lee}},
  \bibinfo {author} {\bibfnamefont {A.}~\bibnamefont {Richardella}}, \bibinfo
  {author} {\bibfnamefont {J.~L.}\ \bibnamefont {Grab}}, \bibinfo {author}
  {\bibfnamefont {P.~J.}\ \bibnamefont {Mintun}}, \bibinfo {author}
  {\bibfnamefont {M.~H.}\ \bibnamefont {Fischer}}, \bibinfo {author}
  {\bibfnamefont {A.}~\bibnamefont {Vaezi}}, \bibinfo {author} {\bibfnamefont
  {A.}~\bibnamefont {Manchon}}, \bibinfo {author} {\bibfnamefont {E.-A.}\
  \bibnamefont {Kim}}, \bibinfo {author} {\bibfnamefont {N.}~\bibnamefont
  {Samarth}}, \ and\ \bibinfo {author} {\bibfnamefont {D.~C.}\ \bibnamefont
  {Ralph}},\ }\href {\doibase 10.1038/nature13534} {\bibfield  {journal}
  {\bibinfo  {journal} {Nature}\ }\textbf {\bibinfo {volume} {511}},\ \bibinfo
  {pages} {449} (\bibinfo {year} {2014})}\BibitemShut {NoStop}%
\bibitem [{\citenamefont {Pi}\ \emph {et~al.}(2010)\citenamefont {Pi},
  \citenamefont {Won~Kim}, \citenamefont {Bae}, \citenamefont {Lee},
  \citenamefont {Cho}, \citenamefont {Kim},\ and\ \citenamefont
  {Seo}}]{Pi2010}%
  \BibitemOpen
  \bibfield  {author} {\bibinfo {author} {\bibfnamefont {U.~H.}\ \bibnamefont
  {Pi}}, \bibinfo {author} {\bibfnamefont {K.}~\bibnamefont {Won~Kim}},
  \bibinfo {author} {\bibfnamefont {J.~Y.}\ \bibnamefont {Bae}}, \bibinfo
  {author} {\bibfnamefont {S.~C.}\ \bibnamefont {Lee}}, \bibinfo {author}
  {\bibfnamefont {Y.~J.}\ \bibnamefont {Cho}}, \bibinfo {author} {\bibfnamefont
  {K.~S.}\ \bibnamefont {Kim}}, \ and\ \bibinfo {author} {\bibfnamefont
  {S.}~\bibnamefont {Seo}},\ }\href {\doibase 10.1063/1.3502596} {\bibfield
  {journal} {\bibinfo  {journal} {Applied Physics Letters}\ }\textbf {\bibinfo
  {volume} {97}},\ \bibinfo {pages} {162507} (\bibinfo {year}
  {2010})}\BibitemShut {NoStop}%
\bibitem [{\citenamefont {Garello}\ \emph {et~al.}(2013)\citenamefont
  {Garello}, \citenamefont {Miron}, \citenamefont {Avci}, \citenamefont
  {Freimuth}, \citenamefont {Mokrousov}, \citenamefont {Bl{\"u}gel},
  \citenamefont {Auffret}, \citenamefont {Boulle}, \citenamefont {Gaudin},\
  and\ \citenamefont {Gambardella}}]{Garello2013}%
  \BibitemOpen
  \bibfield  {author} {\bibinfo {author} {\bibfnamefont {K.}~\bibnamefont
  {Garello}}, \bibinfo {author} {\bibfnamefont {I.~M.}\ \bibnamefont {Miron}},
  \bibinfo {author} {\bibfnamefont {C.~O.}\ \bibnamefont {Avci}}, \bibinfo
  {author} {\bibfnamefont {F.}~\bibnamefont {Freimuth}}, \bibinfo {author}
  {\bibfnamefont {Y.}~\bibnamefont {Mokrousov}}, \bibinfo {author}
  {\bibfnamefont {S.}~\bibnamefont {Bl{\"u}gel}}, \bibinfo {author}
  {\bibfnamefont {S.}~\bibnamefont {Auffret}}, \bibinfo {author} {\bibfnamefont
  {O.}~\bibnamefont {Boulle}}, \bibinfo {author} {\bibfnamefont
  {G.}~\bibnamefont {Gaudin}}, \ and\ \bibinfo {author} {\bibfnamefont
  {P.}~\bibnamefont {Gambardella}},\ }\href {\doibase 10.1038/nnano.2013.145}
  {\bibfield  {journal} {\bibinfo  {journal} {Nature Nanotechnology}\ }\textbf
  {\bibinfo {volume} {8}},\ \bibinfo {pages} {587} (\bibinfo {year}
  {2013})}\BibitemShut {NoStop}%
\bibitem [{\citenamefont {Hayashi}\ \emph {et~al.}(2014)\citenamefont
  {Hayashi}, \citenamefont {Kim}, \citenamefont {Yamanouchi},\ and\
  \citenamefont {Ohno}}]{Hayashi2014}%
  \BibitemOpen
  \bibfield  {author} {\bibinfo {author} {\bibfnamefont {M.}~\bibnamefont
  {Hayashi}}, \bibinfo {author} {\bibfnamefont {J.}~\bibnamefont {Kim}},
  \bibinfo {author} {\bibfnamefont {M.}~\bibnamefont {Yamanouchi}}, \ and\
  \bibinfo {author} {\bibfnamefont {H.}~\bibnamefont {Ohno}},\ }\href {\doibase
  10.1103/PhysRevB.89.144425} {\bibfield  {journal} {\bibinfo  {journal} {Phys.
  Rev. B}\ }\textbf {\bibinfo {volume} {89}},\ \bibinfo {pages} {144425}
  (\bibinfo {year} {2014})}\BibitemShut {NoStop}%
\bibitem [{\citenamefont {Fan}\ \emph {et~al.}(2014)\citenamefont {Fan},
  \citenamefont {Celik}, \citenamefont {Wu}, \citenamefont {Ni}, \citenamefont
  {Lee}, \citenamefont {Lorenz},\ and\ \citenamefont {Xiao}}]{Fan2014}%
  \BibitemOpen
  \bibfield  {author} {\bibinfo {author} {\bibfnamefont {X.}~\bibnamefont
  {Fan}}, \bibinfo {author} {\bibfnamefont {H.}~\bibnamefont {Celik}}, \bibinfo
  {author} {\bibfnamefont {J.}~\bibnamefont {Wu}}, \bibinfo {author}
  {\bibfnamefont {C.}~\bibnamefont {Ni}}, \bibinfo {author} {\bibfnamefont
  {K.-J.}\ \bibnamefont {Lee}}, \bibinfo {author} {\bibfnamefont {V.~O.}\
  \bibnamefont {Lorenz}}, \ and\ \bibinfo {author} {\bibfnamefont {J.~Q.}\
  \bibnamefont {Xiao}},\ }\href {\doibase 10.1038/ncomms4042} {\bibfield
  {journal} {\bibinfo  {journal} {Nature Communications}\ }\textbf {\bibinfo
  {volume} {5}},\ \bibinfo {pages} {3042} (\bibinfo {year} {2014})}\BibitemShut
  {NoStop}%
\bibitem [{\citenamefont {Fan}\ \emph {et~al.}(2016)\citenamefont {Fan},
  \citenamefont {Mellnik}, \citenamefont {Wang}, \citenamefont {Reynolds},
  \citenamefont {Wang}, \citenamefont {Celik}, \citenamefont {Lorenz},
  \citenamefont {Ralph},\ and\ \citenamefont {Xiao}}]{Fan2016}%
  \BibitemOpen
  \bibfield  {author} {\bibinfo {author} {\bibfnamefont {X.}~\bibnamefont
  {Fan}}, \bibinfo {author} {\bibfnamefont {A.~R.}\ \bibnamefont {Mellnik}},
  \bibinfo {author} {\bibfnamefont {W.}~\bibnamefont {Wang}}, \bibinfo {author}
  {\bibfnamefont {N.}~\bibnamefont {Reynolds}}, \bibinfo {author}
  {\bibfnamefont {T.}~\bibnamefont {Wang}}, \bibinfo {author} {\bibfnamefont
  {H.}~\bibnamefont {Celik}}, \bibinfo {author} {\bibfnamefont {V.~O.}\
  \bibnamefont {Lorenz}}, \bibinfo {author} {\bibfnamefont {D.~C.}\
  \bibnamefont {Ralph}}, \ and\ \bibinfo {author} {\bibfnamefont {J.~Q.}\
  \bibnamefont {Xiao}},\ }\href {\doibase 10.1063/1.4962402} {\bibfield
  {journal} {\bibinfo  {journal} {Applied Physics Letters}\ }\textbf {\bibinfo
  {volume} {109}},\ \bibinfo {pages} {122406} (\bibinfo {year}
  {2016})}\BibitemShut {NoStop}%
\bibitem [{\citenamefont {Sun}(2000)}]{Sun2000}%
  \BibitemOpen
  \bibfield  {author} {\bibinfo {author} {\bibfnamefont {J.~Z.}\ \bibnamefont
  {Sun}},\ }\href {\doibase 10.1103/physrevb.62.570} {\bibfield  {journal}
  {\bibinfo  {journal} {Physical Review B}\ }\textbf {\bibinfo {volume} {62}},\
  \bibinfo {pages} {570} (\bibinfo {year} {2000})}\BibitemShut {NoStop}%
\bibitem [{\citenamefont {Althammer}\ \emph {et~al.}(2013)\citenamefont
  {Althammer}, \citenamefont {Meyer}, \citenamefont {Nakayama}, \citenamefont
  {Schreier}, \citenamefont {Altmannshofer}, \citenamefont {Weiler},
  \citenamefont {Huebl}, \citenamefont {Gepr\"ags}, \citenamefont {Opel},
  \citenamefont {Gross}, \citenamefont {Meier}, \citenamefont {Klewe},
  \citenamefont {Kuschel}, \citenamefont {Schmalhorst}, \citenamefont {Reiss},
  \citenamefont {Shen}, \citenamefont {Gupta}, \citenamefont {Chen},
  \citenamefont {Bauer}, \citenamefont {Saitoh},\ and\ \citenamefont
  {Goennenwein}}]{Althammer2013}%
  \BibitemOpen
  \bibfield  {author} {\bibinfo {author} {\bibfnamefont {M.}~\bibnamefont
  {Althammer}}, \bibinfo {author} {\bibfnamefont {S.}~\bibnamefont {Meyer}},
  \bibinfo {author} {\bibfnamefont {H.}~\bibnamefont {Nakayama}}, \bibinfo
  {author} {\bibfnamefont {M.}~\bibnamefont {Schreier}}, \bibinfo {author}
  {\bibfnamefont {S.}~\bibnamefont {Altmannshofer}}, \bibinfo {author}
  {\bibfnamefont {M.}~\bibnamefont {Weiler}}, \bibinfo {author} {\bibfnamefont
  {H.}~\bibnamefont {Huebl}}, \bibinfo {author} {\bibfnamefont
  {S.}~\bibnamefont {Gepr\"ags}}, \bibinfo {author} {\bibfnamefont
  {M.}~\bibnamefont {Opel}}, \bibinfo {author} {\bibfnamefont {R.}~\bibnamefont
  {Gross}}, \bibinfo {author} {\bibfnamefont {D.}~\bibnamefont {Meier}},
  \bibinfo {author} {\bibfnamefont {C.}~\bibnamefont {Klewe}}, \bibinfo
  {author} {\bibfnamefont {T.}~\bibnamefont {Kuschel}}, \bibinfo {author}
  {\bibfnamefont {J.-M.}\ \bibnamefont {Schmalhorst}}, \bibinfo {author}
  {\bibfnamefont {G.}~\bibnamefont {Reiss}}, \bibinfo {author} {\bibfnamefont
  {L.}~\bibnamefont {Shen}}, \bibinfo {author} {\bibfnamefont {A.}~\bibnamefont
  {Gupta}}, \bibinfo {author} {\bibfnamefont {Y.-T.}\ \bibnamefont {Chen}},
  \bibinfo {author} {\bibfnamefont {G.~E.~W.}\ \bibnamefont {Bauer}}, \bibinfo
  {author} {\bibfnamefont {E.}~\bibnamefont {Saitoh}}, \ and\ \bibinfo {author}
  {\bibfnamefont {S.~T.~B.}\ \bibnamefont {Goennenwein}},\ }\href {\doibase
  10.1103/PhysRevB.87.224401} {\bibfield  {journal} {\bibinfo  {journal} {Phys.
  Rev. B}\ }\textbf {\bibinfo {volume} {87}},\ \bibinfo {pages} {224401}
  (\bibinfo {year} {2013})}\BibitemShut {NoStop}%
\bibitem [{\citenamefont {Kim}\ \emph {et~al.}(2016)\citenamefont {Kim},
  \citenamefont {Sheng}, \citenamefont {Takahashi}, \citenamefont {Mitani},\
  and\ \citenamefont {Hayashi}}]{Kim2016}%
  \BibitemOpen
  \bibfield  {author} {\bibinfo {author} {\bibfnamefont {J.}~\bibnamefont
  {Kim}}, \bibinfo {author} {\bibfnamefont {P.}~\bibnamefont {Sheng}}, \bibinfo
  {author} {\bibfnamefont {S.}~\bibnamefont {Takahashi}}, \bibinfo {author}
  {\bibfnamefont {S.}~\bibnamefont {Mitani}}, \ and\ \bibinfo {author}
  {\bibfnamefont {M.}~\bibnamefont {Hayashi}},\ }\href {\doibase
  10.1103/PhysRevLett.116.097201} {\bibfield  {journal} {\bibinfo  {journal}
  {Phys. Rev. Lett.}\ }\textbf {\bibinfo {volume} {116}},\ \bibinfo {pages}
  {097201} (\bibinfo {year} {2016})}\BibitemShut {NoStop}%
\bibitem [{\citenamefont {Emori}\ \emph {et~al.}(2014)\citenamefont {Emori},
  \citenamefont {Martinez}, \citenamefont {Lee}, \citenamefont {Lee},
  \citenamefont {Bauer}, \citenamefont {Ahn}, \citenamefont {Agrawal},
  \citenamefont {Bono},\ and\ \citenamefont {Beach}}]{Emori2014}%
  \BibitemOpen
  \bibfield  {author} {\bibinfo {author} {\bibfnamefont {S.}~\bibnamefont
  {Emori}}, \bibinfo {author} {\bibfnamefont {E.}~\bibnamefont {Martinez}},
  \bibinfo {author} {\bibfnamefont {K.-J.}\ \bibnamefont {Lee}}, \bibinfo
  {author} {\bibfnamefont {H.-W.}\ \bibnamefont {Lee}}, \bibinfo {author}
  {\bibfnamefont {U.}~\bibnamefont {Bauer}}, \bibinfo {author} {\bibfnamefont
  {S.-M.}\ \bibnamefont {Ahn}}, \bibinfo {author} {\bibfnamefont
  {P.}~\bibnamefont {Agrawal}}, \bibinfo {author} {\bibfnamefont {D.~C.}\
  \bibnamefont {Bono}}, \ and\ \bibinfo {author} {\bibfnamefont {G.~S.~D.}\
  \bibnamefont {Beach}},\ }\href {\doibase 10.1103/PhysRevB.90.184427}
  {\bibfield  {journal} {\bibinfo  {journal} {Phys. Rev. B}\ }\textbf {\bibinfo
  {volume} {90}},\ \bibinfo {pages} {184427} (\bibinfo {year}
  {2014})}\BibitemShut {NoStop}%
\bibitem [{\citenamefont {Pai}\ \emph {et~al.}(2016)\citenamefont {Pai},
  \citenamefont {Mann}, \citenamefont {Tan},\ and\ \citenamefont
  {Beach}}]{Pai2016}%
  \BibitemOpen
  \bibfield  {author} {\bibinfo {author} {\bibfnamefont {C.-F.}\ \bibnamefont
  {Pai}}, \bibinfo {author} {\bibfnamefont {M.}~\bibnamefont {Mann}}, \bibinfo
  {author} {\bibfnamefont {A.~J.}\ \bibnamefont {Tan}}, \ and\ \bibinfo
  {author} {\bibfnamefont {G.~S.~D.}\ \bibnamefont {Beach}},\ }\href {\doibase
  10.1103/PhysRevB.93.144409} {\bibfield  {journal} {\bibinfo  {journal} {Phys.
  Rev. B}\ }\textbf {\bibinfo {volume} {93}},\ \bibinfo {pages} {144409}
  (\bibinfo {year} {2016})}\BibitemShut {NoStop}%
\bibitem [{\citenamefont {Pai}\ \emph {et~al.}(2015)\citenamefont {Pai},
  \citenamefont {Ou}, \citenamefont {Vilela-Le\~ao}, \citenamefont {Ralph},\
  and\ \citenamefont {Buhrman}}]{Pai2015}%
  \BibitemOpen
  \bibfield  {author} {\bibinfo {author} {\bibfnamefont {C.-F.}\ \bibnamefont
  {Pai}}, \bibinfo {author} {\bibfnamefont {Y.}~\bibnamefont {Ou}}, \bibinfo
  {author} {\bibfnamefont {L.~H.}\ \bibnamefont {Vilela-Le\~ao}}, \bibinfo
  {author} {\bibfnamefont {D.~C.}\ \bibnamefont {Ralph}}, \ and\ \bibinfo
  {author} {\bibfnamefont {R.~A.}\ \bibnamefont {Buhrman}},\ }\href {\doibase
  10.1103/PhysRevB.92.064426} {\bibfield  {journal} {\bibinfo  {journal} {Phys.
  Rev. B}\ }\textbf {\bibinfo {volume} {92}},\ \bibinfo {pages} {064426}
  (\bibinfo {year} {2015})}\BibitemShut {NoStop}%
\bibitem [{\citenamefont {Tao}\ \emph {et~al.}(2018)\citenamefont {Tao},
  \citenamefont {Liu}, \citenamefont {Miao}, \citenamefont {Yu}, \citenamefont
  {Feng}, \citenamefont {Sun}, \citenamefont {You}, \citenamefont {Du},
  \citenamefont {Chen}, \citenamefont {Zhang}, \citenamefont {Zhang},
  \citenamefont {Yuan}, \citenamefont {Wu},\ and\ \citenamefont
  {Ding}}]{Tao2018}%
  \BibitemOpen
  \bibfield  {author} {\bibinfo {author} {\bibfnamefont {X.}~\bibnamefont
  {Tao}}, \bibinfo {author} {\bibfnamefont {Q.}~\bibnamefont {Liu}}, \bibinfo
  {author} {\bibfnamefont {B.}~\bibnamefont {Miao}}, \bibinfo {author}
  {\bibfnamefont {R.}~\bibnamefont {Yu}}, \bibinfo {author} {\bibfnamefont
  {Z.}~\bibnamefont {Feng}}, \bibinfo {author} {\bibfnamefont {L.}~\bibnamefont
  {Sun}}, \bibinfo {author} {\bibfnamefont {B.}~\bibnamefont {You}}, \bibinfo
  {author} {\bibfnamefont {J.}~\bibnamefont {Du}}, \bibinfo {author}
  {\bibfnamefont {K.}~\bibnamefont {Chen}}, \bibinfo {author} {\bibfnamefont
  {S.}~\bibnamefont {Zhang}}, \bibinfo {author} {\bibfnamefont
  {L.}~\bibnamefont {Zhang}}, \bibinfo {author} {\bibfnamefont
  {Z.}~\bibnamefont {Yuan}}, \bibinfo {author} {\bibfnamefont {D.}~\bibnamefont
  {Wu}}, \ and\ \bibinfo {author} {\bibfnamefont {H.}~\bibnamefont {Ding}},\
  }\href {\doibase 10.1126/sciadv.aat1670} {\bibfield  {journal} {\bibinfo
  {journal} {Science Advances}\ }\textbf {\bibinfo {volume} {4}},\ \bibinfo
  {pages} {1670} (\bibinfo {year} {2018})}\BibitemShut {NoStop}%
\bibitem [{\citenamefont {Zhu}\ \emph {et~al.}(2019{\natexlab{a}})\citenamefont
  {Zhu}, \citenamefont {Sobotkiewich}, \citenamefont {Ma}, \citenamefont {Li},
  \citenamefont {Ralph},\ and\ \citenamefont {Buhrman}}]{Zhu2019AFM}%
  \BibitemOpen
  \bibfield  {author} {\bibinfo {author} {\bibfnamefont {L.}~\bibnamefont
  {Zhu}}, \bibinfo {author} {\bibfnamefont {K.}~\bibnamefont {Sobotkiewich}},
  \bibinfo {author} {\bibfnamefont {X.}~\bibnamefont {Ma}}, \bibinfo {author}
  {\bibfnamefont {X.}~\bibnamefont {Li}}, \bibinfo {author} {\bibfnamefont
  {D.~C.}\ \bibnamefont {Ralph}}, \ and\ \bibinfo {author} {\bibfnamefont
  {R.~A.}\ \bibnamefont {Buhrman}},\ }\href {\doibase 10.1002/adfm.201805822}
  {\bibfield  {journal} {\bibinfo  {journal} {Advanced Functional Materials}\
  }\textbf {\bibinfo {volume} {29}},\ \bibinfo {pages} {1805822} (\bibinfo
  {year} {2019}{\natexlab{a}})}\BibitemShut {NoStop}%
\bibitem [{\citenamefont {Lau}\ and\ \citenamefont {Hayashi}(2017)}]{Lau2017}%
  \BibitemOpen
  \bibfield  {author} {\bibinfo {author} {\bibfnamefont {Y.-C.}\ \bibnamefont
  {Lau}}\ and\ \bibinfo {author} {\bibfnamefont {M.}~\bibnamefont {Hayashi}},\
  }\href {\doibase 10.7567/jjap.56.0802b5} {\bibfield  {journal} {\bibinfo
  {journal} {Japanese Journal of Applied Physics}\ }\textbf {\bibinfo {volume}
  {56}},\ \bibinfo {pages} {0802B5} (\bibinfo {year} {2017})}\BibitemShut
  {NoStop}%
\bibitem [{\citenamefont {Tserkovnyak}\ \emph
  {et~al.}(2002{\natexlab{a}})\citenamefont {Tserkovnyak}, \citenamefont
  {Brataas},\ and\ \citenamefont {Bauer}}]{Tserkovnyak2002}%
  \BibitemOpen
  \bibfield  {author} {\bibinfo {author} {\bibfnamefont {Y.}~\bibnamefont
  {Tserkovnyak}}, \bibinfo {author} {\bibfnamefont {A.}~\bibnamefont
  {Brataas}}, \ and\ \bibinfo {author} {\bibfnamefont {G.~E.~W.}\ \bibnamefont
  {Bauer}},\ }\href {\doibase 10.1103/PhysRevLett.88.117601} {\bibfield
  {journal} {\bibinfo  {journal} {Phys. Rev. Lett.}\ }\textbf {\bibinfo
  {volume} {88}},\ \bibinfo {pages} {117601} (\bibinfo {year}
  {2002}{\natexlab{a}})}\BibitemShut {NoStop}%
\bibitem [{\citenamefont {Tserkovnyak}\ \emph
  {et~al.}(2002{\natexlab{b}})\citenamefont {Tserkovnyak}, \citenamefont
  {Brataas},\ and\ \citenamefont {Bauer}}]{Tserkovnyak2002a}%
  \BibitemOpen
  \bibfield  {author} {\bibinfo {author} {\bibfnamefont {Y.}~\bibnamefont
  {Tserkovnyak}}, \bibinfo {author} {\bibfnamefont {A.}~\bibnamefont
  {Brataas}}, \ and\ \bibinfo {author} {\bibfnamefont {G.~E.~W.}\ \bibnamefont
  {Bauer}},\ }\href {\doibase 10.1103/PhysRevB.66.224403} {\bibfield  {journal}
  {\bibinfo  {journal} {Phys. Rev. B}\ }\textbf {\bibinfo {volume} {66}},\
  \bibinfo {pages} {224403} (\bibinfo {year} {2002}{\natexlab{b}})}\BibitemShut
  {NoStop}%
\bibitem [{\citenamefont {Mosendz}\ \emph {et~al.}(2010)\citenamefont
  {Mosendz}, \citenamefont {Vlaminck}, \citenamefont {Pearson}, \citenamefont
  {Fradin}, \citenamefont {Bauer}, \citenamefont {Bader},\ and\ \citenamefont
  {Hoffmann}}]{Mosendz2010a}%
  \BibitemOpen
  \bibfield  {author} {\bibinfo {author} {\bibfnamefont {O.}~\bibnamefont
  {Mosendz}}, \bibinfo {author} {\bibfnamefont {V.}~\bibnamefont {Vlaminck}},
  \bibinfo {author} {\bibfnamefont {J.~E.}\ \bibnamefont {Pearson}}, \bibinfo
  {author} {\bibfnamefont {F.~Y.}\ \bibnamefont {Fradin}}, \bibinfo {author}
  {\bibfnamefont {G.~E.~W.}\ \bibnamefont {Bauer}}, \bibinfo {author}
  {\bibfnamefont {S.~D.}\ \bibnamefont {Bader}}, \ and\ \bibinfo {author}
  {\bibfnamefont {A.}~\bibnamefont {Hoffmann}},\ }\href {\doibase
  10.1103/PhysRevB.82.214403} {\bibfield  {journal} {\bibinfo  {journal} {Phys.
  Rev. B}\ }\textbf {\bibinfo {volume} {82}},\ \bibinfo {pages} {214403}
  (\bibinfo {year} {2010})}\BibitemShut {NoStop}%
\bibitem [{\citenamefont {Azevedo}\ \emph {et~al.}(2011)\citenamefont
  {Azevedo}, \citenamefont {Vilela-Le\~ao}, \citenamefont
  {Rodr\'{\i}guez-Su\'arez}, \citenamefont {Lacerda~Santos},\ and\
  \citenamefont {Rezende}}]{Azevedo2011}%
  \BibitemOpen
  \bibfield  {author} {\bibinfo {author} {\bibfnamefont {A.}~\bibnamefont
  {Azevedo}}, \bibinfo {author} {\bibfnamefont {L.~H.}\ \bibnamefont
  {Vilela-Le\~ao}}, \bibinfo {author} {\bibfnamefont {R.~L.}\ \bibnamefont
  {Rodr\'{\i}guez-Su\'arez}}, \bibinfo {author} {\bibfnamefont {A.~F.}\
  \bibnamefont {Lacerda~Santos}}, \ and\ \bibinfo {author} {\bibfnamefont
  {S.~M.}\ \bibnamefont {Rezende}},\ }\href {\doibase
  10.1103/PhysRevB.83.144402} {\bibfield  {journal} {\bibinfo  {journal} {Phys.
  Rev. B}\ }\textbf {\bibinfo {volume} {83}},\ \bibinfo {pages} {144402}
  (\bibinfo {year} {2011})}\BibitemShut {NoStop}%
\bibitem [{\citenamefont {Uchida}\ \emph {et~al.}(2010)\citenamefont {Uchida},
  \citenamefont {Adachi}, \citenamefont {Ota}, \citenamefont {Nakayama},
  \citenamefont {Maekawa},\ and\ \citenamefont {Saitoh}}]{Uchida2010}%
  \BibitemOpen
  \bibfield  {author} {\bibinfo {author} {\bibfnamefont {K.-i.}\ \bibnamefont
  {Uchida}}, \bibinfo {author} {\bibfnamefont {H.}~\bibnamefont {Adachi}},
  \bibinfo {author} {\bibfnamefont {T.}~\bibnamefont {Ota}}, \bibinfo {author}
  {\bibfnamefont {H.}~\bibnamefont {Nakayama}}, \bibinfo {author}
  {\bibfnamefont {S.}~\bibnamefont {Maekawa}}, \ and\ \bibinfo {author}
  {\bibfnamefont {E.}~\bibnamefont {Saitoh}},\ }\href {\doibase
  10.1063/1.3507386} {\bibfield  {journal} {\bibinfo  {journal} {Applied
  Physics Letters}\ }\textbf {\bibinfo {volume} {97}},\ \bibinfo {pages}
  {172505} (\bibinfo {year} {2010})}\BibitemShut {NoStop}%
\bibitem [{\citenamefont {Holanda}\ \emph {et~al.}(2017)\citenamefont
  {Holanda}, \citenamefont {Alves~Santos}, \citenamefont
  {Rodr\'{\i}guez-Su\'arez}, \citenamefont {Azevedo},\ and\ \citenamefont
  {Rezende}}]{Holanda2017}%
  \BibitemOpen
  \bibfield  {author} {\bibinfo {author} {\bibfnamefont {J.}~\bibnamefont
  {Holanda}}, \bibinfo {author} {\bibfnamefont {O.}~\bibnamefont
  {Alves~Santos}}, \bibinfo {author} {\bibfnamefont {R.~L.}\ \bibnamefont
  {Rodr\'{\i}guez-Su\'arez}}, \bibinfo {author} {\bibfnamefont
  {A.}~\bibnamefont {Azevedo}}, \ and\ \bibinfo {author} {\bibfnamefont
  {S.~M.}\ \bibnamefont {Rezende}},\ }\href {\doibase
  10.1103/PhysRevB.95.134432} {\bibfield  {journal} {\bibinfo  {journal} {Phys.
  Rev. B}\ }\textbf {\bibinfo {volume} {95}},\ \bibinfo {pages} {134432}
  (\bibinfo {year} {2017})}\BibitemShut {NoStop}%
\bibitem [{\citenamefont {Lee}\ \emph {et~al.}(2015)\citenamefont {Lee},
  \citenamefont {Kim}, \citenamefont {Yeon~Lee}, \citenamefont {Kim},
  \citenamefont {Lee}, \citenamefont {Lee}, \citenamefont {Jeong},
  \citenamefont {Lee}, \citenamefont {Song}, \citenamefont {Sohn},
  \citenamefont {Shin},\ and\ \citenamefont {Park}}]{Lee2015}%
  \BibitemOpen
  \bibfield  {author} {\bibinfo {author} {\bibfnamefont {K.-D.}\ \bibnamefont
  {Lee}}, \bibinfo {author} {\bibfnamefont {D.-J.}\ \bibnamefont {Kim}},
  \bibinfo {author} {\bibfnamefont {H.}~\bibnamefont {Yeon~Lee}}, \bibinfo
  {author} {\bibfnamefont {S.-H.}\ \bibnamefont {Kim}}, \bibinfo {author}
  {\bibfnamefont {J.-H.}\ \bibnamefont {Lee}}, \bibinfo {author} {\bibfnamefont
  {K.-M.}\ \bibnamefont {Lee}}, \bibinfo {author} {\bibfnamefont {J.-R.}\
  \bibnamefont {Jeong}}, \bibinfo {author} {\bibfnamefont {K.-S.}\ \bibnamefont
  {Lee}}, \bibinfo {author} {\bibfnamefont {H.-S.}\ \bibnamefont {Song}},
  \bibinfo {author} {\bibfnamefont {J.-W.}\ \bibnamefont {Sohn}}, \bibinfo
  {author} {\bibfnamefont {S.-C.}\ \bibnamefont {Shin}}, \ and\ \bibinfo
  {author} {\bibfnamefont {B.-G.}\ \bibnamefont {Park}},\ }\href {\doibase
  10.1038/srep10249} {\bibfield  {journal} {\bibinfo  {journal} {Scientific
  Reports}\ }\textbf {\bibinfo {volume} {5}},\ \bibinfo {pages} {10249}
  (\bibinfo {year} {2015})}\BibitemShut {NoStop}%
\bibitem [{\citenamefont {Kikkawa}\ \emph {et~al.}(2013)\citenamefont
  {Kikkawa}, \citenamefont {Uchida}, \citenamefont {Shiomi}, \citenamefont
  {Qiu}, \citenamefont {Hou}, \citenamefont {Tian}, \citenamefont {Nakayama},
  \citenamefont {Jin},\ and\ \citenamefont {Saitoh}}]{kikkawa2013}%
  \BibitemOpen
  \bibfield  {author} {\bibinfo {author} {\bibfnamefont {T.}~\bibnamefont
  {Kikkawa}}, \bibinfo {author} {\bibfnamefont {K.}~\bibnamefont {Uchida}},
  \bibinfo {author} {\bibfnamefont {Y.}~\bibnamefont {Shiomi}}, \bibinfo
  {author} {\bibfnamefont {Z.}~\bibnamefont {Qiu}}, \bibinfo {author}
  {\bibfnamefont {D.}~\bibnamefont {Hou}}, \bibinfo {author} {\bibfnamefont
  {D.}~\bibnamefont {Tian}}, \bibinfo {author} {\bibfnamefont {H.}~\bibnamefont
  {Nakayama}}, \bibinfo {author} {\bibfnamefont {X.-F.}\ \bibnamefont {Jin}}, \
  and\ \bibinfo {author} {\bibfnamefont {E.}~\bibnamefont {Saitoh}},\ }\href
  {\doibase 10.1103/PhysRevLett.110.067207} {\bibfield  {journal} {\bibinfo
  {journal} {Phys. Rev. Lett.}\ }\textbf {\bibinfo {volume} {110}},\ \bibinfo
  {pages} {067207} (\bibinfo {year} {2013})}\BibitemShut {NoStop}%
\bibitem [{\citenamefont {Avci}\ \emph {et~al.}(2014)\citenamefont {Avci},
  \citenamefont {Garello}, \citenamefont {Gabureac}, \citenamefont {Ghosh},
  \citenamefont {Fuhrer}, \citenamefont {Alvarado},\ and\ \citenamefont
  {Gambardella}}]{Avci2014}%
  \BibitemOpen
  \bibfield  {author} {\bibinfo {author} {\bibfnamefont {C.~O.}\ \bibnamefont
  {Avci}}, \bibinfo {author} {\bibfnamefont {K.}~\bibnamefont {Garello}},
  \bibinfo {author} {\bibfnamefont {M.}~\bibnamefont {Gabureac}}, \bibinfo
  {author} {\bibfnamefont {A.}~\bibnamefont {Ghosh}}, \bibinfo {author}
  {\bibfnamefont {A.}~\bibnamefont {Fuhrer}}, \bibinfo {author} {\bibfnamefont
  {S.~F.}\ \bibnamefont {Alvarado}}, \ and\ \bibinfo {author} {\bibfnamefont
  {P.}~\bibnamefont {Gambardella}},\ }\href {\doibase
  10.1103/PhysRevB.90.224427} {\bibfield  {journal} {\bibinfo  {journal} {Phys.
  Rev. B}\ }\textbf {\bibinfo {volume} {90}},\ \bibinfo {pages} {224427}
  (\bibinfo {year} {2014})}\BibitemShut {NoStop}%
\bibitem [{\citenamefont {Roschewsky}\ \emph {et~al.}(2019)\citenamefont
  {Roschewsky}, \citenamefont {Walker}, \citenamefont {Gowtham}, \citenamefont
  {Muschinske}, \citenamefont {Hellman}, \citenamefont {Bank},\ and\
  \citenamefont {Salahuddin}}]{Roschewsky2019}%
  \BibitemOpen
  \bibfield  {author} {\bibinfo {author} {\bibfnamefont {N.}~\bibnamefont
  {Roschewsky}}, \bibinfo {author} {\bibfnamefont {E.~S.}\ \bibnamefont
  {Walker}}, \bibinfo {author} {\bibfnamefont {P.}~\bibnamefont {Gowtham}},
  \bibinfo {author} {\bibfnamefont {S.}~\bibnamefont {Muschinske}}, \bibinfo
  {author} {\bibfnamefont {F.}~\bibnamefont {Hellman}}, \bibinfo {author}
  {\bibfnamefont {S.~R.}\ \bibnamefont {Bank}}, \ and\ \bibinfo {author}
  {\bibfnamefont {S.}~\bibnamefont {Salahuddin}},\ }\href {\doibase
  10.1103/PhysRevB.99.195103} {\bibfield  {journal} {\bibinfo  {journal} {Phys.
  Rev. B}\ }\textbf {\bibinfo {volume} {99}},\ \bibinfo {pages} {195103}
  (\bibinfo {year} {2019})}\BibitemShut {NoStop}%
\bibitem [{\citenamefont {Kondou}\ \emph {et~al.}(2016)\citenamefont {Kondou},
  \citenamefont {Sukegawa}, \citenamefont {Kasai}, \citenamefont {Mitani},
  \citenamefont {Niimi},\ and\ \citenamefont {Otani}}]{Kondou2016}%
  \BibitemOpen
  \bibfield  {author} {\bibinfo {author} {\bibfnamefont {K.}~\bibnamefont
  {Kondou}}, \bibinfo {author} {\bibfnamefont {H.}~\bibnamefont {Sukegawa}},
  \bibinfo {author} {\bibfnamefont {S.}~\bibnamefont {Kasai}}, \bibinfo
  {author} {\bibfnamefont {S.}~\bibnamefont {Mitani}}, \bibinfo {author}
  {\bibfnamefont {Y.}~\bibnamefont {Niimi}}, \ and\ \bibinfo {author}
  {\bibfnamefont {Y.}~\bibnamefont {Otani}},\ }\href {\doibase
  10.7567/apex.9.023002} {\bibfield  {journal} {\bibinfo  {journal} {Applied
  Physics Express}\ }\textbf {\bibinfo {volume} {9}},\ \bibinfo {pages}
  {023002} (\bibinfo {year} {2016})}\BibitemShut {NoStop}%
\bibitem [{\citenamefont {Okada}\ \emph {et~al.}(2019)\citenamefont {Okada},
  \citenamefont {Takeuchi}, \citenamefont {Furuya}, \citenamefont {Zhang},
  \citenamefont {Sato}, \citenamefont {Fukami},\ and\ \citenamefont
  {Ohno}}]{Okada2019}%
  \BibitemOpen
  \bibfield  {author} {\bibinfo {author} {\bibfnamefont {A.}~\bibnamefont
  {Okada}}, \bibinfo {author} {\bibfnamefont {Y.}~\bibnamefont {Takeuchi}},
  \bibinfo {author} {\bibfnamefont {K.}~\bibnamefont {Furuya}}, \bibinfo
  {author} {\bibfnamefont {C.}~\bibnamefont {Zhang}}, \bibinfo {author}
  {\bibfnamefont {H.}~\bibnamefont {Sato}}, \bibinfo {author} {\bibfnamefont
  {S.}~\bibnamefont {Fukami}}, \ and\ \bibinfo {author} {\bibfnamefont
  {H.}~\bibnamefont {Ohno}},\ }\href {\doibase
  10.1103/PhysRevApplied.12.014040} {\bibfield  {journal} {\bibinfo  {journal}
  {Phys. Rev. Applied}\ }\textbf {\bibinfo {volume} {12}},\ \bibinfo {pages}
  {014040} (\bibinfo {year} {2019})}\BibitemShut {NoStop}%
\bibitem [{\citenamefont {Kumar}\ \emph {et~al.}(2017)\citenamefont {Kumar},
  \citenamefont {Akansel}, \citenamefont {Stopfel}, \citenamefont {Fazlali},
  \citenamefont {\AA{}kerman}, \citenamefont {Brucas},\ and\ \citenamefont
  {Svedlindh}}]{Kumar2019}%
  \BibitemOpen
  \bibfield  {author} {\bibinfo {author} {\bibfnamefont {A.}~\bibnamefont
  {Kumar}}, \bibinfo {author} {\bibfnamefont {S.}~\bibnamefont {Akansel}},
  \bibinfo {author} {\bibfnamefont {H.}~\bibnamefont {Stopfel}}, \bibinfo
  {author} {\bibfnamefont {M.}~\bibnamefont {Fazlali}}, \bibinfo {author}
  {\bibfnamefont {J.}~\bibnamefont {\AA{}kerman}}, \bibinfo {author}
  {\bibfnamefont {R.}~\bibnamefont {Brucas}}, \ and\ \bibinfo {author}
  {\bibfnamefont {P.}~\bibnamefont {Svedlindh}},\ }\href {\doibase
  10.1103/PhysRevB.95.064406} {\bibfield  {journal} {\bibinfo  {journal} {Phys.
  Rev. B}\ }\textbf {\bibinfo {volume} {95}},\ \bibinfo {pages} {064406}
  (\bibinfo {year} {2017})}\BibitemShut {NoStop}%
\bibitem [{\citenamefont {Tulapurkar}\ \emph {et~al.}(2005)\citenamefont
  {Tulapurkar}, \citenamefont {Suzuki}, \citenamefont {Fukushima},
  \citenamefont {Kubota}, \citenamefont {Maehara}, \citenamefont {Tsunekawa},
  \citenamefont {Djayaprawira}, \citenamefont {Watanabe},\ and\ \citenamefont
  {Yuasa}}]{Tulapurkar2005}%
  \BibitemOpen
  \bibfield  {author} {\bibinfo {author} {\bibfnamefont {A.}~\bibnamefont
  {Tulapurkar}}, \bibinfo {author} {\bibfnamefont {Y.}~\bibnamefont {Suzuki}},
  \bibinfo {author} {\bibfnamefont {A.}~\bibnamefont {Fukushima}}, \bibinfo
  {author} {\bibfnamefont {H.}~\bibnamefont {Kubota}}, \bibinfo {author}
  {\bibfnamefont {H.}~\bibnamefont {Maehara}}, \bibinfo {author} {\bibfnamefont
  {K.}~\bibnamefont {Tsunekawa}}, \bibinfo {author} {\bibfnamefont
  {D.}~\bibnamefont {Djayaprawira}}, \bibinfo {author} {\bibfnamefont
  {N.}~\bibnamefont {Watanabe}}, \ and\ \bibinfo {author} {\bibfnamefont
  {S.}~\bibnamefont {Yuasa}},\ }\href {\doibase 10.1038/nature04207} {\bibfield
   {journal} {\bibinfo  {journal} {Nature}\ }\textbf {\bibinfo {volume}
  {438}},\ \bibinfo {pages} {339} (\bibinfo {year} {2005})}\BibitemShut
  {NoStop}%
\bibitem [{\citenamefont {Sankey}\ \emph {et~al.}(2006)\citenamefont {Sankey},
  \citenamefont {Braganca}, \citenamefont {Garcia}, \citenamefont {Krivorotov},
  \citenamefont {Buhrman},\ and\ \citenamefont {Ralph}}]{Sankey2006}%
  \BibitemOpen
  \bibfield  {author} {\bibinfo {author} {\bibfnamefont {J.~C.}\ \bibnamefont
  {Sankey}}, \bibinfo {author} {\bibfnamefont {P.~M.}\ \bibnamefont
  {Braganca}}, \bibinfo {author} {\bibfnamefont {A.~G.~F.}\ \bibnamefont
  {Garcia}}, \bibinfo {author} {\bibfnamefont {I.~N.}\ \bibnamefont
  {Krivorotov}}, \bibinfo {author} {\bibfnamefont {R.~A.}\ \bibnamefont
  {Buhrman}}, \ and\ \bibinfo {author} {\bibfnamefont {D.~C.}\ \bibnamefont
  {Ralph}},\ }\href {\doibase 10.1103/PhysRevLett.96.227601} {\bibfield
  {journal} {\bibinfo  {journal} {Phys. Rev. Lett.}\ }\textbf {\bibinfo
  {volume} {96}},\ \bibinfo {pages} {227601} (\bibinfo {year}
  {2006})}\BibitemShut {NoStop}%
\bibitem [{\citenamefont {Lustikova}\ \emph {et~al.}(2015)\citenamefont
  {Lustikova}, \citenamefont {Shiomi},\ and\ \citenamefont
  {Saitoh}}]{Lustikova2015}%
  \BibitemOpen
  \bibfield  {author} {\bibinfo {author} {\bibfnamefont {J.}~\bibnamefont
  {Lustikova}}, \bibinfo {author} {\bibfnamefont {Y.}~\bibnamefont {Shiomi}}, \
  and\ \bibinfo {author} {\bibfnamefont {E.}~\bibnamefont {Saitoh}},\ }\href
  {\doibase 10.1103/PhysRevB.92.224436} {\bibfield  {journal} {\bibinfo
  {journal} {Phys. Rev. B}\ }\textbf {\bibinfo {volume} {92}},\ \bibinfo
  {pages} {224436} (\bibinfo {year} {2015})}\BibitemShut {NoStop}%
\bibitem [{\citenamefont {Jungfleisch}\ \emph {et~al.}(2015)\citenamefont
  {Jungfleisch}, \citenamefont {Chumak}, \citenamefont {Kehlberger},
  \citenamefont {Lauer}, \citenamefont {Kim}, \citenamefont {Onbasli},
  \citenamefont {Ross}, \citenamefont {Kl\"aui},\ and\ \citenamefont
  {Hillebrands}}]{jungfleisch2015}%
  \BibitemOpen
  \bibfield  {author} {\bibinfo {author} {\bibfnamefont {M.~B.}\ \bibnamefont
  {Jungfleisch}}, \bibinfo {author} {\bibfnamefont {A.~V.}\ \bibnamefont
  {Chumak}}, \bibinfo {author} {\bibfnamefont {A.}~\bibnamefont {Kehlberger}},
  \bibinfo {author} {\bibfnamefont {V.}~\bibnamefont {Lauer}}, \bibinfo
  {author} {\bibfnamefont {D.~H.}\ \bibnamefont {Kim}}, \bibinfo {author}
  {\bibfnamefont {M.~C.}\ \bibnamefont {Onbasli}}, \bibinfo {author}
  {\bibfnamefont {C.~A.}\ \bibnamefont {Ross}}, \bibinfo {author}
  {\bibfnamefont {M.}~\bibnamefont {Kl\"aui}}, \ and\ \bibinfo {author}
  {\bibfnamefont {B.}~\bibnamefont {Hillebrands}},\ }\href {\doibase
  10.1103/PhysRevB.91.134407} {\bibfield  {journal} {\bibinfo  {journal} {Phys.
  Rev. B}\ }\textbf {\bibinfo {volume} {91}},\ \bibinfo {pages} {134407}
  (\bibinfo {year} {2015})}\BibitemShut {NoStop}%
\bibitem [{\citenamefont {Nakayama}\ \emph {et~al.}(2012)\citenamefont
  {Nakayama}, \citenamefont {Ando}, \citenamefont {Harii}, \citenamefont
  {Yoshino}, \citenamefont {Takahashi}, \citenamefont {Kajiwara}, \citenamefont
  {Uchida}, \citenamefont {Fujikawa},\ and\ \citenamefont
  {Saitoh}}]{Nakayama2012}%
  \BibitemOpen
  \bibfield  {author} {\bibinfo {author} {\bibfnamefont {H.}~\bibnamefont
  {Nakayama}}, \bibinfo {author} {\bibfnamefont {K.}~\bibnamefont {Ando}},
  \bibinfo {author} {\bibfnamefont {K.}~\bibnamefont {Harii}}, \bibinfo
  {author} {\bibfnamefont {T.}~\bibnamefont {Yoshino}}, \bibinfo {author}
  {\bibfnamefont {R.}~\bibnamefont {Takahashi}}, \bibinfo {author}
  {\bibfnamefont {Y.}~\bibnamefont {Kajiwara}}, \bibinfo {author}
  {\bibfnamefont {K.}~\bibnamefont {Uchida}}, \bibinfo {author} {\bibfnamefont
  {Y.}~\bibnamefont {Fujikawa}}, \ and\ \bibinfo {author} {\bibfnamefont
  {E.}~\bibnamefont {Saitoh}},\ }\href {\doibase 10.1103/PhysRevB.85.144408}
  {\bibfield  {journal} {\bibinfo  {journal} {Phys. Rev. B}\ }\textbf {\bibinfo
  {volume} {85}},\ \bibinfo {pages} {144408} (\bibinfo {year}
  {2012})}\BibitemShut {NoStop}%
\bibitem [{\citenamefont {Rezende}\ \emph {et~al.}(2014)\citenamefont
  {Rezende}, \citenamefont {Rodr\'{\i}guez-Su\'arez}, \citenamefont {Cunha},
  \citenamefont {Rodrigues}, \citenamefont {Machado}, \citenamefont
  {Fonseca~Guerra}, \citenamefont {Lopez~Ortiz},\ and\ \citenamefont
  {Azevedo}}]{Rezende2014}%
  \BibitemOpen
  \bibfield  {author} {\bibinfo {author} {\bibfnamefont {S.~M.}\ \bibnamefont
  {Rezende}}, \bibinfo {author} {\bibfnamefont {R.~L.}\ \bibnamefont
  {Rodr\'{\i}guez-Su\'arez}}, \bibinfo {author} {\bibfnamefont {R.~O.}\
  \bibnamefont {Cunha}}, \bibinfo {author} {\bibfnamefont {A.~R.}\ \bibnamefont
  {Rodrigues}}, \bibinfo {author} {\bibfnamefont {F.~L.~A.}\ \bibnamefont
  {Machado}}, \bibinfo {author} {\bibfnamefont {G.~A.}\ \bibnamefont
  {Fonseca~Guerra}}, \bibinfo {author} {\bibfnamefont {J.~C.}\ \bibnamefont
  {Lopez~Ortiz}}, \ and\ \bibinfo {author} {\bibfnamefont {A.}~\bibnamefont
  {Azevedo}},\ }\href {\doibase 10.1103/PhysRevB.89.014416} {\bibfield
  {journal} {\bibinfo  {journal} {Phys. Rev. B}\ }\textbf {\bibinfo {volume}
  {89}},\ \bibinfo {pages} {014416} (\bibinfo {year} {2014})}\BibitemShut
  {NoStop}%
\bibitem [{\citenamefont {Saitoh}\ \emph {et~al.}(2006)\citenamefont {Saitoh},
  \citenamefont {Ueda}, \citenamefont {Miyajima},\ and\ \citenamefont
  {Tatara}}]{Saitoh2006}%
  \BibitemOpen
  \bibfield  {author} {\bibinfo {author} {\bibfnamefont {E.}~\bibnamefont
  {Saitoh}}, \bibinfo {author} {\bibfnamefont {M.}~\bibnamefont {Ueda}},
  \bibinfo {author} {\bibfnamefont {H.}~\bibnamefont {Miyajima}}, \ and\
  \bibinfo {author} {\bibfnamefont {G.}~\bibnamefont {Tatara}},\ }\href
  {\doibase 10.1063/1.2199473} {\bibfield  {journal} {\bibinfo  {journal}
  {Applied Physics Letters}\ }\textbf {\bibinfo {volume} {88}},\ \bibinfo
  {pages} {182509} (\bibinfo {year} {2006})}\BibitemShut {NoStop}%
\bibitem [{\citenamefont {Iguchi}\ and\ \citenamefont
  {Saitoh}(2017)}]{Iguchi2017}%
  \BibitemOpen
  \bibfield  {author} {\bibinfo {author} {\bibfnamefont {R.}~\bibnamefont
  {Iguchi}}\ and\ \bibinfo {author} {\bibfnamefont {E.}~\bibnamefont
  {Saitoh}},\ }\href {\doibase 10.7566/JPSJ.86.011003} {\bibfield  {journal}
  {\bibinfo  {journal} {Journal of the Physical Society of Japan}\ }\textbf
  {\bibinfo {volume} {86}},\ \bibinfo {pages} {011003} (\bibinfo {year}
  {2017})}\BibitemShut {NoStop}%
\bibitem [{\citenamefont {Bose}\ \emph {et~al.}(2017)\citenamefont {Bose},
  \citenamefont {Dutta}, \citenamefont {Bhuktare}, \citenamefont {Singh},\ and\
  \citenamefont {Tulapurkar}}]{bose2017}%
  \BibitemOpen
  \bibfield  {author} {\bibinfo {author} {\bibfnamefont {A.}~\bibnamefont
  {Bose}}, \bibinfo {author} {\bibfnamefont {S.}~\bibnamefont {Dutta}},
  \bibinfo {author} {\bibfnamefont {S.}~\bibnamefont {Bhuktare}}, \bibinfo
  {author} {\bibfnamefont {H.}~\bibnamefont {Singh}}, \ and\ \bibinfo {author}
  {\bibfnamefont {A.~A.}\ \bibnamefont {Tulapurkar}},\ }\href {\doibase
  10.1063/1.4999948} {\bibfield  {journal} {\bibinfo  {journal} {Applied
  Physics Letters}\ }\textbf {\bibinfo {volume} {111}},\ \bibinfo {pages}
  {162405} (\bibinfo {year} {2017})}\BibitemShut {NoStop}%
\bibitem [{\citenamefont {Keller}\ \emph {et~al.}(2017)\citenamefont {Keller},
  \citenamefont {Greser}, \citenamefont {Schweizer}, \citenamefont {Conca},
  \citenamefont {Lauer}, \citenamefont {Dubs}, \citenamefont {Hillebrands},\
  and\ \citenamefont {Papaioannou}}]{Keller2017}%
  \BibitemOpen
  \bibfield  {author} {\bibinfo {author} {\bibfnamefont {S.}~\bibnamefont
  {Keller}}, \bibinfo {author} {\bibfnamefont {J.}~\bibnamefont {Greser}},
  \bibinfo {author} {\bibfnamefont {M.~R.}\ \bibnamefont {Schweizer}}, \bibinfo
  {author} {\bibfnamefont {A.}~\bibnamefont {Conca}}, \bibinfo {author}
  {\bibfnamefont {V.}~\bibnamefont {Lauer}}, \bibinfo {author} {\bibfnamefont
  {C.}~\bibnamefont {Dubs}}, \bibinfo {author} {\bibfnamefont {B.}~\bibnamefont
  {Hillebrands}}, \ and\ \bibinfo {author} {\bibfnamefont {E.~T.}\ \bibnamefont
  {Papaioannou}},\ }\href {\doibase 10.1103/physrevb.96.024437} {\bibfield
  {journal} {\bibinfo  {journal} {Physical Review B}\ }\textbf {\bibinfo
  {volume} {96}} (\bibinfo {year} {2017}),\
  10.1103/physrevb.96.024437}\BibitemShut {NoStop}%
\bibitem [{\citenamefont {Harder}\ \emph {et~al.}(2016)\citenamefont {Harder},
  \citenamefont {Gui},\ and\ \citenamefont {Hu}}]{harder2016}%
  \BibitemOpen
  \bibfield  {author} {\bibinfo {author} {\bibfnamefont {M.}~\bibnamefont
  {Harder}}, \bibinfo {author} {\bibfnamefont {Y.}~\bibnamefont {Gui}}, \ and\
  \bibinfo {author} {\bibfnamefont {C.-M.}\ \bibnamefont {Hu}},\ }\href
  {\doibase https://doi.org/10.1016/j.physrep.2016.10.002} {\bibfield
  {journal} {\bibinfo  {journal} {Physics Reports}\ }\textbf {\bibinfo {volume}
  {661}},\ \bibinfo {pages} {1 } (\bibinfo {year} {2016})},\ \bibinfo {note}
  {electrical detection of magnetization dynamics via spin rectification
  effects}\BibitemShut {NoStop}%
\bibitem [{Sup()}]{Supplement}%
  \BibitemOpen
  \href@noop {} {\bibinfo  {journal} {Supplemental Materials: Transverse and
  Longitudinal Spin-Torque Ferromagnetic Resonance for Improved Measurements of
  Spin-Orbit Torques}\ }\BibitemShut {NoStop}%
\bibitem [{\citenamefont {Schreier}\ \emph {et~al.}(2014)\citenamefont
  {Schreier}, \citenamefont {Bauer}, \citenamefont {Vasyuchka}, \citenamefont
  {Flipse}, \citenamefont {ichi Uchida}, \citenamefont {Lotze}, \citenamefont
  {Lauer}, \citenamefont {Chumak}, \citenamefont {Serga}, \citenamefont
  {Daimon}, \citenamefont {Kikkawa}, \citenamefont {Saitoh}, \citenamefont {van
  Wees}, \citenamefont {Hillebrands}, \citenamefont {Gross},\ and\
  \citenamefont {Goennenwein}}]{schreier2014}%
  \BibitemOpen
\bibfield  {journal} {  }\bibfield  {author} {\bibinfo {author} {\bibfnamefont
  {M.}~\bibnamefont {Schreier}}, \bibinfo {author} {\bibfnamefont {G.~E.~W.}\
  \bibnamefont {Bauer}}, \bibinfo {author} {\bibfnamefont {V.~I.}\ \bibnamefont
  {Vasyuchka}}, \bibinfo {author} {\bibfnamefont {J.}~\bibnamefont {Flipse}},
  \bibinfo {author} {\bibfnamefont {K.}~\bibnamefont {ichi Uchida}}, \bibinfo
  {author} {\bibfnamefont {J.}~\bibnamefont {Lotze}}, \bibinfo {author}
  {\bibfnamefont {V.}~\bibnamefont {Lauer}}, \bibinfo {author} {\bibfnamefont
  {A.~V.}\ \bibnamefont {Chumak}}, \bibinfo {author} {\bibfnamefont {A.~A.}\
  \bibnamefont {Serga}}, \bibinfo {author} {\bibfnamefont {S.}~\bibnamefont
  {Daimon}}, \bibinfo {author} {\bibfnamefont {T.}~\bibnamefont {Kikkawa}},
  \bibinfo {author} {\bibfnamefont {E.}~\bibnamefont {Saitoh}}, \bibinfo
  {author} {\bibfnamefont {B.~J.}\ \bibnamefont {van Wees}}, \bibinfo {author}
  {\bibfnamefont {B.}~\bibnamefont {Hillebrands}}, \bibinfo {author}
  {\bibfnamefont {R.}~\bibnamefont {Gross}}, \ and\ \bibinfo {author}
  {\bibfnamefont {S.~T.~B.}\ \bibnamefont {Goennenwein}},\ }\href {\doibase
  10.1088/0022-3727/48/2/025001} {\bibfield  {journal} {\bibinfo  {journal}
  {Journal of Physics D: Applied Physics}\ }\textbf {\bibinfo {volume} {48}},\
  \bibinfo {pages} {025001} (\bibinfo {year} {2014})}\BibitemShut {NoStop}%
\bibitem [{\citenamefont {Fan}\ \emph {et~al.}(2013)\citenamefont {Fan},
  \citenamefont {Wu}, \citenamefont {Chen}, \citenamefont {Jerry},
  \citenamefont {Zhang},\ and\ \citenamefont {Xiao}}]{fan2013}%
  \BibitemOpen
  \bibfield  {author} {\bibinfo {author} {\bibfnamefont {X.}~\bibnamefont
  {Fan}}, \bibinfo {author} {\bibfnamefont {J.}~\bibnamefont {Wu}}, \bibinfo
  {author} {\bibfnamefont {Y.}~\bibnamefont {Chen}}, \bibinfo {author}
  {\bibfnamefont {M.~J.}\ \bibnamefont {Jerry}}, \bibinfo {author}
  {\bibfnamefont {H.}~\bibnamefont {Zhang}}, \ and\ \bibinfo {author}
  {\bibfnamefont {J.~Q.}\ \bibnamefont {Xiao}},\ }\href {\doibase
  10.1038/ncomms2709} {\bibfield  {journal} {\bibinfo  {journal} {Nature
  Communications}\ }\textbf {\bibinfo {volume} {4}},\ \bibinfo {pages} {1799}
  (\bibinfo {year} {2013})}\BibitemShut {NoStop}%
\bibitem [{\citenamefont {Nan}\ \emph {et~al.}(2015)\citenamefont {Nan},
  \citenamefont {Emori}, \citenamefont {Boone}, \citenamefont {Wang},
  \citenamefont {Oxholm}, \citenamefont {Jones}, \citenamefont {Howe},
  \citenamefont {Brown},\ and\ \citenamefont {Sun}}]{nan2015}%
  \BibitemOpen
  \bibfield  {author} {\bibinfo {author} {\bibfnamefont {T.}~\bibnamefont
  {Nan}}, \bibinfo {author} {\bibfnamefont {S.}~\bibnamefont {Emori}}, \bibinfo
  {author} {\bibfnamefont {C.~T.}\ \bibnamefont {Boone}}, \bibinfo {author}
  {\bibfnamefont {X.}~\bibnamefont {Wang}}, \bibinfo {author} {\bibfnamefont
  {T.~M.}\ \bibnamefont {Oxholm}}, \bibinfo {author} {\bibfnamefont {J.~G.}\
  \bibnamefont {Jones}}, \bibinfo {author} {\bibfnamefont {B.~M.}\ \bibnamefont
  {Howe}}, \bibinfo {author} {\bibfnamefont {G.~J.}\ \bibnamefont {Brown}}, \
  and\ \bibinfo {author} {\bibfnamefont {N.~X.}\ \bibnamefont {Sun}},\ }\href
  {\doibase 10.1103/PhysRevB.91.214416} {\bibfield  {journal} {\bibinfo
  {journal} {Phys. Rev. B}\ }\textbf {\bibinfo {volume} {91}},\ \bibinfo
  {pages} {214416} (\bibinfo {year} {2015})}\BibitemShut {NoStop}%
\bibitem [{\citenamefont {Zhu}\ \emph {et~al.}(2019{\natexlab{b}})\citenamefont
  {Zhu}, \citenamefont {Ralph},\ and\ \citenamefont {Buhrman}}]{Zhu2019geff}%
  \BibitemOpen
  \bibfield  {author} {\bibinfo {author} {\bibfnamefont {L.}~\bibnamefont
  {Zhu}}, \bibinfo {author} {\bibfnamefont {D.~C.}\ \bibnamefont {Ralph}}, \
  and\ \bibinfo {author} {\bibfnamefont {R.~A.}\ \bibnamefont {Buhrman}},\
  }\href {\doibase 10.1103/PhysRevLett.123.057203} {\bibfield  {journal}
  {\bibinfo  {journal} {Phys. Rev. Lett.}\ }\textbf {\bibinfo {volume} {123}},\
  \bibinfo {pages} {057203} (\bibinfo {year} {2019}{\natexlab{b}})}\BibitemShut
  {NoStop}%
\bibitem [{\citenamefont {Nguyen}\ \emph {et~al.}(2016)\citenamefont {Nguyen},
  \citenamefont {Ralph},\ and\ \citenamefont {Buhrman}}]{Nguyen2016}%
  \BibitemOpen
  \bibfield  {author} {\bibinfo {author} {\bibfnamefont {M.-H.}\ \bibnamefont
  {Nguyen}}, \bibinfo {author} {\bibfnamefont {D.~C.}\ \bibnamefont {Ralph}}, \
  and\ \bibinfo {author} {\bibfnamefont {R.~A.}\ \bibnamefont {Buhrman}},\
  }\href {\doibase 10.1103/PhysRevLett.116.126601} {\bibfield  {journal}
  {\bibinfo  {journal} {Phys. Rev. Lett.}\ }\textbf {\bibinfo {volume} {116}},\
  \bibinfo {pages} {126601} (\bibinfo {year} {2016})}\BibitemShut {NoStop}%
\bibitem [{\citenamefont {Iguchi}\ \emph {et~al.}(2012)\citenamefont {Iguchi},
  \citenamefont {Ando}, \citenamefont {Takahashi}, \citenamefont {An},
  \citenamefont {Saitoh},\ and\ \citenamefont {Sato}}]{Iguchi2012}%
  \BibitemOpen
  \bibfield  {author} {\bibinfo {author} {\bibfnamefont {R.}~\bibnamefont
  {Iguchi}}, \bibinfo {author} {\bibfnamefont {K.}~\bibnamefont {Ando}},
  \bibinfo {author} {\bibfnamefont {R.}~\bibnamefont {Takahashi}}, \bibinfo
  {author} {\bibfnamefont {T.}~\bibnamefont {An}}, \bibinfo {author}
  {\bibfnamefont {E.}~\bibnamefont {Saitoh}}, \ and\ \bibinfo {author}
  {\bibfnamefont {T.}~\bibnamefont {Sato}},\ }\href {\doibase
  10.1143/jjap.51.103004} {\bibfield  {journal} {\bibinfo  {journal} {Japanese
  Journal of Applied Physics}\ }\textbf {\bibinfo {volume} {51}},\ \bibinfo
  {pages} {103004} (\bibinfo {year} {2012})}\BibitemShut {NoStop}%
\end{thebibliography}%



\section*{Supplementary material}

# Supplementary Information:

# Transverse and Longitudinal Spin-Torque Ferromagnetic Resonance for Improved Measurements of Spin-Orbit Torques

Saba Karimeddiny,[1, *] Joseph A. Mittelstaedt,[1, *] Robert A. Buhrman,[1] and Daniel C. Ralph[1, 2]

[1]*Cornell University, Ithaca, NY 14850, USA*

[2]*Kavli Institute at Cornell, Ithaca, NY 14853, USA*

(Dated: July 3, 2020)

**CONTENTS**

---

[*] These two authors contributed equally

# I. DEFINITIONS OF COORDINATE AXES

It will be convenient to define two coordinate systems. Capital letters ($\hat{X},\hat{Y},\hat{Z}$) will denote a coordinate system fixed with respect to the device structure. $\hat{X}$ is the direction of microwave current flow, $\hat{Z}$ is perpendicular to the sample plane, and $\hat{Y} = \hat{Z} \times \hat{X}$. Lower-case letters ($\hat{x},\hat{y},\hat{z}$) will denote a coordinate system defined relative to the precession axis of the magnetization. $\hat{y}$ is along the precession axis, $\hat{z} = \hat{Z}$ with $z = 0$ corresponding the the HM/FM interface, and $\hat{x} = \hat{y} \times \hat{z}$. Spherical polar coordinates will also be used to specify the magnetization direction, with the polar angle $\theta$ measured with respect to $\hat{Z}$ and the azimuthal angle $\phi$ measured with respect to $\hat{X}$.

# II. SOLVING THE LLGS EQUATION FOR THE OSCILLATORY COMPONENTS OF THE MAGNETIZATION

We wish to compute the dynamics of the magnetization in a heavy metal (HM)/ferromagnet (FM) bilayer in response to a microwave current applied within the sample plane. Since by our convention the precession axis lies along $\hat{y}$, in response to the microwave current the oscillatory components of the magnetization are $m_x$ and $m_z$. We begin with Landau-Lifshitz-Gilbert-Slonczewski (LLGS) equation [1, 2].

$$\dot{\hat{m}} = \alpha \hat{m} \times \dot{\hat{m}} + \vec{\tau}_{\text{neq}} + \vec{\tau}_{\text{eq}} \tag{1}$$

where $\hat{m}$ is the magnetization orientation, $\alpha$ is the Gilbert damping parameter, and $\vec{\tau}_{\text{neq}}$ is the current-induced non-equilibrium spin-orbit torque (SOT). $\vec{\tau}_{\text{eq}} = -\gamma \left( \hat{m} \times -\frac{d\widetilde{F}}{d\hat{m}} \right)$ is the equilibrium torque ($\gamma = 2\mu_B/\hbar$ is the gyromagnetic ratio with $\mu_B$ the Bohr magneton), which can be found from the magnetic free energy density, $\widetilde{F}$. We may write the magnetic free energy density as

$$\widetilde{F} = F/M_s = -\vec{B} \cdot \hat{m} + \frac{1}{2}\mu_0 M_s \hat{m} \cdot \overleftrightarrow{N} \cdot \hat{m} - \frac{K_\perp}{M_s}(\hat{m} \cdot \hat{n})^2 - \frac{K_\parallel}{M_s}(\hat{m} \cdot \hat{u})^2 \tag{2}$$

where $M_s$ is the saturation magnetization, $\vec{B}$ is the applied magnetic field, $\overleftrightarrow{N}$ is the demagnetization tensor, $K_\perp(K_\parallel)$ is the strength of the out-of-plane (within-plane) anisotropy, $\hat{n}$ is the film normal and $\hat{u}$ is the in-plane anisotropy direction. Our polycrystalline samples have negligible anisotropy within the plane so we neglect the final term, and we use the thin-film approximation that the demagnetization tensor has only one nonzero element $N_{ZZ} = 1$. Our equilibrium torque is therefore

$$\vec{\tau}_{\text{eq}} = \gamma \begin{pmatrix} m_y m_z \mu_0 M_{\text{eff}} + m_z B \\ -m_x m_z \mu_0 M_{\text{eff}} \\ -m_x B \end{pmatrix}. \tag{3}$$

Here $\mu_0 M_{\text{eff}} = \mu_0 M_s - 2K_\perp/M_s$. We are working in coordinates where $m_y \approx 1$ and $m_x, m_z \ll 1$ so to leading order

$$\vec{\tau}_{\text{eq}} = \begin{pmatrix} m_z \omega_2 \\ 0 \\ -m_x \omega_1 \end{pmatrix}, \tag{4}$$

where we have defined $\omega_1 = \gamma B$ and $\omega_2 = \gamma(B + \mu_0 M_{\text{eff}})$.

The damping term is of the form, to leading order

$$\alpha \hat{m} \times \dot{\hat{m}} = \alpha \begin{pmatrix} m_y \dot{m}_z - m_z \dot{m}_y \\ m_z \dot{m}_x - m_x \dot{m}_z \\ m_x \dot{m}_y - m_y \dot{m}_x \end{pmatrix} \approx \alpha \begin{pmatrix} \dot{m}_z \\ 0 \\ -\dot{m}_x \end{pmatrix}. \tag{5}$$

The amplitudes of the oscillatory non-equilibrium torques have the form

$$\vec{\tau}_{\text{neq}} = \begin{pmatrix} \tau_x \\ 0 \\ \tau_z \end{pmatrix}. \tag{6}$$

We will assume that the torques are in-phase with the applied current and arise from a spin current with the symmetry required for the spin Hall effect in polycrystalline materials:

$$\begin{aligned} \tau_x &\sim (\hat{m} \times \hat{\sigma} \times \hat{m}) = -\hat{m} \times (\hat{Y} \times \hat{m}) \\ \tau_z &\sim (\hat{\sigma} \times \hat{m}) = -\hat{Y} \times \hat{m}, \end{aligned} \tag{7}$$

where for Pt the spin Hall effect orients the spin moment $\hat{\sigma} = -\hat{Y}$ for current flowing along $+\hat{X}$. We can write this more explicitly:

$$\begin{aligned} \tau_x &= \tau_{\text{DL}}^0 \cos(\phi) \\ \tau_z &= \tau_z^0 \cos(\phi) = (\tau_{\text{FL}}^0 + \tau_{\text{Oe}}^0) \cos(\phi). \end{aligned} \tag{8}$$

Here we have defined

$$\begin{aligned} \tau_{\text{DL}}^0 &= \xi_{\text{DL}} \frac{\mu_B J_e}{e M_S t_{\text{FM}}} \\ \tau_{\text{FL}}^0 &= \xi_{\text{FL}} \frac{\mu_B J_e}{e M_S t_{\text{FM}}} \\ \tau_{\text{Oe}}^0 &= \frac{\mu_0 \gamma J_e t_{\text{HM}}}{2}, \end{aligned} \tag{9}$$

where $\xi_{\text{DL}}$ is the damping-like spin-orbit torque efficiency, $\xi_{\text{FL}}$ is the field-like spin-orbit torque efficiency, $J_e$ is the charge current density in the heavy metal, $e$ is the electron charge, $t_{\text{FM}}$ is the

thickness of the ferromagnet layer, and $t_{\text{HM}}$ is the thickness of the heavy metal. The damping-like torque efficiency $\xi_{\text{DL}}$ will have a value reduced from the intrinsic spin Hall ratio within the heavy metal ($\theta_{\text{SH}}$) by the spin-transparency factor of the HM/FM interface.

Assembling Equations (4)-(6), the linearized LLGS equation we aim to solve takes the form

$$\begin{pmatrix} \dot{m}_x \\ 0 \\ \dot{m}_z \end{pmatrix} = \begin{pmatrix} \alpha \dot{m}_z + m_z \omega_2 + \tau_x \\ 0 \\ -\alpha \dot{m}_x - m_x \omega_1 + \tau_z \end{pmatrix}. \tag{10}$$

We assume that we will have an oscillatory solution and so use the ansatz $m_{x(z)}(t) = m_{x(z)} e^{-i\omega t}$. With this substitution, the LLGS equation becomes

$$\begin{pmatrix} -i\omega m_x \\ 0 \\ -i\omega m_z \end{pmatrix} = \begin{pmatrix} -m_z(i\omega\alpha - \omega_2) + \tau_x \\ 0 \\ m_x(i\omega\alpha - \omega_1) + \tau_z \end{pmatrix}, \tag{11}$$

with the solution

$$m_x = \frac{-\omega_2 \tau_z + i\omega \tau_x}{(\omega^2 - \omega_0^2) + i\omega\alpha\omega^+} \tag{12}$$

$$m_z = \frac{\omega_1 \tau_x + i\omega \tau_z}{(\omega^2 - \omega_0^2) + i\omega\alpha\omega^+}. \tag{13}$$

We have defined $\omega_0^2 = \omega_1 \omega_2$ and $\omega^+ = \omega_1 + \omega_2$, and have made an assumption that $\alpha$ is small so that $\tau_x + \alpha\tau_z \approx \tau_x$ and $\tau_z + \alpha\tau_x \approx \tau_z$.

In ST-FMR measurements, we usually perform field sweeps instead of frequency sweeps. To convert our expression, we need to expand about the resonant field:

$$\omega_0^2 \approx \omega_0^2|_{B_0} + (B - B_0) \left.\frac{d(\omega_0^2)}{dB}\right|_{B_0}, \tag{14}$$

where the derivative is

$$\left.\frac{d(\omega_0^2)}{dB}\right|_{B_0} = \omega_{2,B_0} \left.\frac{d\omega_1}{dB}\right|_{B_0} + \omega_{1,B_0} \left.\frac{d\omega_2}{dB}\right|_{B_0} = \gamma \omega_{B_0}^+. \tag{15}$$

The $B_0$ subscript indicates that those frequencies are evaluated at the resonant field, so that $\omega_0^2|_{B_0} = \omega^2$.

$$\omega^2 - \omega_0^2 \approx -\gamma(B - B_0)\omega_{B_0}^+ \tag{16}$$

and hence

$$m_x = \frac{-\omega_2 \tau_z + i\omega \tau_x}{-\gamma(B - B_0)\omega^+ + i\omega\alpha\omega^+} \tag{17}$$

$$m_z = \frac{\omega_1 \tau_x + i\omega \tau_z}{-\gamma(B - B_0)\omega^+ + i\omega\alpha\omega^+}. \tag{18}$$

Near resonance, we can evaluate these expressions using $\omega_1 = \gamma B_0$, $\omega_2 = \gamma(B_0 + \mu_0 M_\text{eff})$, and $\omega^+ = \gamma(2B_0 + \mu_0 M_\text{eff})$, as in the main text.

## III. MAGNETORESISTANCES AND RECTIFICATION IN THE LONGITUDINAL MEASUREMENT GEOMETRY

The total magnetoresistance for a longitudinal measurement can be written in spherical coordinates as

$$R_{XX} = R_0 + R_\text{AMR} \sin^2\theta \cos^2\phi - R_\text{SMR} \sin^2\theta \sin^2\phi. \tag{19}$$

where $R_0$ is a constant offset, $R_\text{AMR}$ is the scale of the anisotropic magnetoresistance, and $R_\text{SMR}$ is the scale of the spin Hall magnetoresistance [3]. We consider small angle precession such that $\theta = \theta_0 + \Delta\theta$ and $\phi = \phi_0 + \Delta\phi$ with $\Delta\theta, \Delta\phi \ll 1$ and expand to get

$$\begin{aligned} R_{XX} = R_0 &+ R_\text{AMR} \left( \sin^2\theta_0 \cos^2\phi_0 + \Delta\theta \sin 2\theta_0 \cos^2\phi_0 - \Delta\phi \sin^2\theta_0 \sin 2\phi_0 \right) \\ &- R_\text{SMR} \left( \sin^2\theta_0 \sin^2\phi_0 + \Delta\theta \sin 2\theta_0 \sin^2\phi_0 + \Delta\phi \sin^2\theta_0 \sin 2\phi_0 \right). \end{aligned} \tag{20}$$

The only pieces of Eq. (20) that are current-rectifiable (able to produce a mixing voltage with the rf current) are the terms linear in $\Delta\theta$ and $\Delta\phi$. For an in-plane magnet we have $\theta_0 = \pi/2$, and therefore the mixing voltage becomes

$$V_{XX}^\text{mix} = \frac{I_\text{RF}}{2}(R_\text{AMR} + R_\text{SMR})(-\Delta\phi \sin 2\phi_0) = \frac{I_\text{RF}}{2}(R_\text{AMR} + R_\text{SMR})(\text{Re}[m_x] \sin 2\phi_0). \tag{21}$$

Only the in-plane deflections of the magnet are rectified to produce a mixing voltage, and the AMR and SMR contributions simply add. For simplicity of notation in the main text, we therefore incorporate both the AMR and SMR contributions in one magnetoresistance amplitude $R_\text{AMR}$.

## IV. SPIN-PUMPING CONTRIBUTION

The precessing magnetization at FMR causes the ferromagnetic layer to inject a spin current into the heavy metal layer; this can create a voltage through the ISHE. The time-averaged spin current in the heavy-metal layer can be written as [4–6]

$$\overleftrightarrow{j}^\text{SP}_{\hat{\sigma}}(z) = \hat{\sigma} \otimes \vec{j}^\text{SP}(z) = \frac{\hbar}{4\pi} g_\text{eff}^{\uparrow\downarrow} \frac{\sinh[(t_\text{HM} + z)/\lambda_\text{sd}]}{\sinh[t_\text{HM}/\lambda_\text{sd}]} \left\langle \hat{m} \times \dot{\hat{m}} \right\rangle \otimes (-\hat{z}) \tag{22}$$

where $\vec{j}^\text{SP} \propto -\hat{z}$ is the direction of the spin current flow, $\hat{\sigma} \propto \left|\left\langle \hat{m} \times \dot{\hat{m}} \right\rangle\right| \parallel -\hat{m}$ is the polarization of the pumped spin current (where the negative sign is to account for *enhanced* Gilbert Damping

due to Spin pumping [7, 8]), $g_{\text{eff}}^{\uparrow\downarrow}$ is the effective spin mixing conductance at the interface, and $\lambda_{\text{sd}}$ is the spin diffusion length of the HM. The resultant voltage is

$$V_{\text{SP}} = -R_{\text{tot}} I = -R_{\text{tot}} \int_{\Sigma_{\text{HM}}} \vec{j}_e^{\text{HM}} \cdot d\vec{A} \tag{23}$$

where $R_{\text{tot}}$ is the total device resistance (that will differ for the longitudinal and transverse cases), $\Sigma_{\text{HM}}$ is the cross-section of the heavy-metal layer, $\vec{j}_e^{\text{HM}} = (2e/\hbar)\theta_{\text{SH}} \vec{j}^{\text{SP}} \times \hat{\sigma}$ is the charge current arising from the ISHE [6] and $d\vec{A}$ is a differential surface area normal, which points along the vector connecting the leads that we are measuring across. The negative sign in Eq. (23) is due to the fact that we are measuring the electric field that arises from the open circuit condition of the device [6]. Simplifying the integrals we have (for the longitudinal geometry)

$$V_{\text{SP}} = -R_{\text{tot}} \int_{\Sigma_{\text{HM}}} \vec{j}_e^{\text{HM}} \cdot d\vec{A} \tag{24}$$

$$= -R_{\text{tot}} \int_{-W/2}^{W/2} \int_0^{-t_{\text{HM}}} \vec{j}_e^{\text{HM}} \cdot d\vec{A} \tag{25}$$

$$= -R_{\text{tot}} W \sin\phi \int_0^{-t_{\text{HM}}} \left|\vec{j}_e^{\text{HM}}\right| dz \tag{26}$$

where W is the width of the Hall bar (dimension along $\hat{Y}$). Note that in the transverse measurement $W\sin\phi \to L\cos\phi$ where $L$ is the device bar length. The only part of $\vec{j}_e^{\text{HM}}$ that depends on the thickness is

$$\int_0^{-t_{\text{HM}}} \frac{\sinh((t_{\text{HM}}+z)/\lambda_{\text{sd}})}{\sinh(t_{\text{HM}}/\lambda_{\text{sd}})} \, dz = \lambda_{\text{sd}} \tanh\left(\frac{t_{\text{HM}}}{2\lambda_{\text{sd}}}\right). \tag{27}$$

At this point, we have (for the longitudinal geometry)

$$V_{\text{SP}} = -\frac{2e}{\hbar}\theta_{\text{SH}} R_{\text{tot}} W \sin\phi \frac{\hbar}{4\pi} g_{\text{eff}}^{\uparrow\downarrow} \lambda_{\text{sd}} \left|\left\langle \hat{m} \times \dot{\hat{m}}\right\rangle\right| \tanh\left(\frac{t_{\text{HM}}}{2\lambda_{\text{sd}}}\right). \tag{28}$$

We now only need to calculate $\left\langle \hat{m} \times \dot{\hat{m}}\right\rangle$, but we already have the oscillatory magnetization components from Sec. II. We can write $\left\langle \hat{m} \times \dot{\hat{m}}\right\rangle = \omega \text{Im}\left[m_x m_z^*\right](-\hat{m})$, so therefore

$$\omega \text{Im}\left[m_x m_z^*\right] = \frac{\omega^2}{(\gamma\omega^+)^2(B-B_0)^2 + (\omega\alpha\omega^+)^2}\left[\omega_1 \tau_x^2 + \omega_2 \tau_z^2\right] \tag{29}$$

$$= \frac{\omega_1 \tau_x^2 + \omega_2 \tau_z^2}{(\alpha\omega^+)^2} S(B), \tag{30}$$

where $S(B) = \Delta^2/\left[(B-B_0)^2 + \Delta^2\right]$ is a symmetric Lorentzian and $\Delta \equiv \omega\alpha/\gamma$. The voltage in the device resulting from the pumped spin can then be written as (for the longitudinal geometry)

$$V_{\text{SP}} = -\frac{2e}{\hbar}\theta_{\text{SH}} R_{\text{tot}} W \sin\phi \frac{\hbar}{4\pi} g_{\text{eff}}^{\uparrow\downarrow} \lambda_{\text{sd}} \left[\frac{\omega_1 \tau_x^2 + \omega_2 \tau_z^2}{(\alpha\omega^+)^2} S(B)\right] \tanh\left(\frac{t_{\text{HM}}}{2\lambda_{\text{sd}}}\right). \tag{31}$$

Putting all of this together, using that $\omega_1 = \gamma B_0$, $\omega_2 = \gamma(B_0+\mu_0 M_{\text{eff}})$, and $\omega^+ = \gamma(2B_0+\mu_0 M_{\text{eff}})$, noting that $\tau_x$ and $\tau_z$ have the angular dependence specified in Eq. (8), and that for the transverse case one has $W\sin\phi \to L\cos\phi$, the spin-pumping voltages in the longitudinal and transverse directions become

$$V_{\text{SP}} = -\frac{eB_0 R_{\text{tot}} \theta_{\text{SH}}}{2\pi\alpha^2 \gamma (2B_0 + \mu_0 M_{\text{eff}})^2} g_{\text{eff}}^{\uparrow\downarrow} \lambda_{\text{sd}} \tanh\left(\frac{t_{\text{HM}}}{2\lambda_{\text{sd}}}\right) \left[(\tau_{\text{SH}}^0)^2 + \left(1 + \frac{\mu_0 M_{\text{eff}}}{B_0}\right)(\tau_{\text{z}}^0)^2\right]$$
$$\times S(B)\cos^2\phi \begin{cases} W\sin\phi, & \text{longitudinal} \\ L\cos\phi, & \text{transverse} \end{cases} \tag{32}$$

$$= -\frac{eB_0 \theta_{\text{SH}}}{2\pi\alpha^2 \gamma (2B_0 + \mu_0 M_{\text{eff}})^2} \frac{1}{\sum_i t_i \sigma_i} g_{\text{eff}}^{\uparrow\downarrow} \lambda_{\text{sd}} \tanh\left(\frac{t_{\text{HM}}}{2\lambda_{\text{sd}}}\right) \left[(\tau_{\text{SH}}^0)^2 + \left(1 + \frac{\mu_0 M_{\text{eff}}}{B_0}\right)(\tau_{\text{z}}^0)^2\right]$$
$$\times S(B)\cos^2\phi \begin{cases} L\sin\phi, & \text{longitudinal} \\ W\cos\phi, & \text{transverse} \end{cases}. \tag{33}$$

In the final equation we have expressed the values of $R_{\text{tot}}$ for the longitudinal and transverse cases in terms of the conductivities of the $i = \text{HM}$ and FM layers added as parallel conductors.

## V. ENERGY ABSORPTION DURING MAGNETIC RESONANCE

We begin with the magnetic free energy per unit area (see section II) assuming no in-plane anisotropy

$$F/A = -\vec{B}\cdot M + \frac{1}{2}\mu_0 t_{\text{FM}} M_s M_{\text{eff}} m_z^2. \tag{34}$$

We assume the external field saturates the magnetization in the y-direction

$$F/A = -Bm_y M_s t_{\text{FM}} + \frac{1}{2}\mu_0 t_{\text{FM}} M_s M_{\text{eff}} m_z^2 \tag{35}$$

and using $|m| = 1$,

$$F/A = -BM_s t_{\text{FM}} + \frac{M_s t_{\text{FM}}}{2}\left[Bm_x^2 + (B+\mu_0 M_{\text{eff}})m_z^2\right]. \tag{36}$$

Taking a time derivative, we have

$$\partial_t F/A = M_s t_{\text{FM}}\left[Bm_x \frac{dm_x}{dt} + (B+\mu_0 M_{\text{eff}})m_z \frac{dm_z}{dt}\right]. \tag{37}$$

To calculate the energy absorbed from the current-induced torques, we set $dm_x/dt = \tau_x$ and $dm_z/dt = \tau_z$. Averaging over one precession cycle, the power per unit area absorbed by the magnet

is

$$\langle \partial_t F/A \rangle = \frac{M_s t_{\text{FM}}}{2} \{B\tau_x \text{Re}[m_x] + (B + \mu_0 M_{\text{eff}})\tau_z \text{Re}[m_z]\} \quad (38)$$

$$= \frac{M_s t_{\text{FM}} \alpha \omega^+}{2\gamma} \left[ \frac{\omega_1 \tau_x^2 + \omega_2 \tau_z^2}{(\alpha \omega^+)^2} \right] S(B). \quad (39)$$

The in-plane torque $\tau_x$ contains contributions from both the antidamping spin-orbit torque and the out-of-plane component of the Oersted field, but when averaged over the width of the sample the antidamping spin-orbit torque gives the larger contribution. Using Eq. (39), with parameter values determined as described in the main text, we have calculated the power absorbed per unit area within the magnetic layer of the Pt/CoFeB samples as a function of the magnetic layer thickness. This is plotted as a fraction of the Ohmic dissipation in the magnetic layer in supplementary Fig. 1. The relative amount of heating for the thinnest samples is greater primarily because of increased magnetic damping for the thinnest samples.

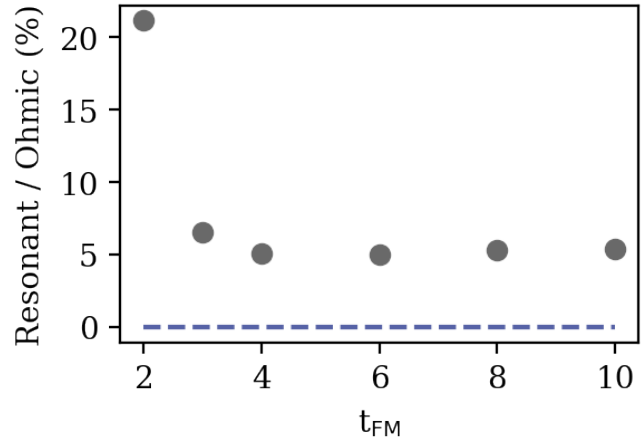

FIG. 1: The ratio of resonant power absorbed to the Ohmic dissipation in the ferromagnetic layer as a function of the ferromagnetic layer thickness, for the Pt(6 nm)/CoFeB($t_{\text{FM}}$) series of samples.

## VI. DEPENDENCE OF THE ARTIFACT VOLTAGE ON RF POWER

Both of the artifact effects discussed in the previous two sections (spin pumping and heating) depend quadratically on the spin torque excitations of the ferromagnet and should therefore depend linearly on the applied RF power. In supplementary Fig. 2 below we show the artifact voltage from the longitudinal ST-FMR signal of a Pt(6)/CoFeB(10) heterostructure. Only one set of points is shown as the average of the AHE/PHE and AHE/AMR correction methods.

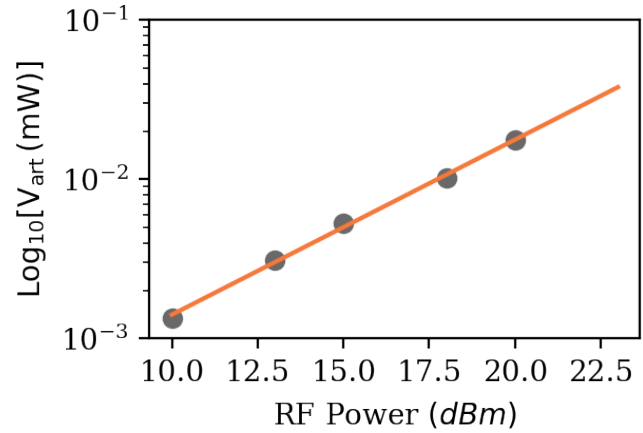

FIG. 2: The logarithm of the artifact voltage vs. the applied RF power in dBm.

The artifact voltage is indeed linearly proportional to the applied power in the regime that we have measured. We note that all measurements in the main text at were performed at 20 dBm.

## VII. CHARACTERIZATION BY VIBRATING SAMPLE MAGNETOMETRY

In this section we show the results of room temperature vibrating sample magnetometry (VSM), which we use to determine the saturation magnetization, $M_s$, and the magnetic dead layer thickness, $t_d$ for each set of ferromagnetic layers. We measure VSM hysteresis loops (not shown) for each thickness of FM that was grown, extract the saturation magnetic moment per unit area for each, and plot the results as in supplementary Fig. 3. We determine $M_s$ of the FM from the slope of data and $t_d$ from the $x$-intercept.

For the $Co_{40}Fe_{40}B_{20}$ sample series, we find $M_s = 1.233(31)$ T and $t_d = 0.056(25)$ nm. The FM thicknesses used in Fig. 3 and Fig. 5 in the main text are adjusted for the dead layer thickness; i.e., $t_{FM} = t_{FM}^{nominal} - t_d$.

## VIII. LONGITUDINAL AND TRANSVERSE ST-FMR DATA FOR OTHER FERROMAGNETS

The results of longitudinal and transverse ST-FMR measurements with different ferromagnet materials are shown in supplementary figures 4 and 5: for a Pt(6 nm)/$Co_{90}Fe_{10}$(6 nm) sample in supplementary Fig. 4 and a Pt(6 nm)/$Ni_{81}Fe_{19}$(8 nm) sample in supplementary Fig. 5. The results

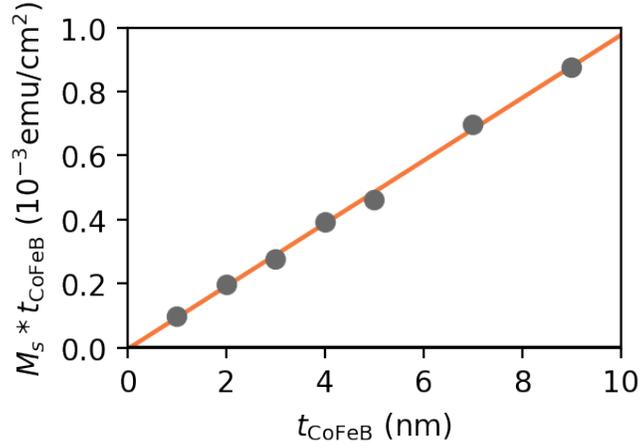

FIG. 3: Saturation Magnetization per unit area vs. the nominal thickness for sputtered $Co_{40}Fe_{40}B_{20}$ layers, all on 6 nm of Pt.

are similar to the CoFeB samples discussed in the main text, except that to obtain good fits for the angular dependence of $S_{XY}$ component requires an additional term approximately proportional to $\sin 2\phi$:

$$S_{XY} = S_{XY}^{\text{PHE}} \cos 2\phi \cos \phi + S_{XY}^{\text{AHE}} \cos \phi + \underbrace{S_{XY}^{2\phi} \sin 2\phi}.$$

The $\sin 2\phi$ term could also include additional angular dependence proportional to even powers of $\cos \phi$ or $\sin \phi$; such variations are difficult to distinguish in the fits. No extra term is needed for the fits to $A_{XY}$ or the longitudinal signals $S_{XX}$ and $A_{XX}$. A contribution $S_{XY} \propto \sin 2\phi$ reflects a difference of overall signal magnitudes between the magnetic field angles $\phi$ and $-\phi$, which (because magnetic field is a pseudovector) can occur only if there is a breaking of mirror symmetry relative to the sample's $\hat{Y}$-$\hat{Z}$ plane. We speculate that the breaking of symmetry is caused by the different contact geometries at the two end of our sample wire, which might cause a longitudinal thermal gradient during resonant heating, and an associated transverse voltage signal with the symmetry of the planar Nernst effect ($\propto m_X m_Y$). As a separate Fourier component, whether or not the $\sin 2\phi$ is included in the fits does not affect the extraction of parameters analyzed in the main text. Table 1 shows how the size of the $\sin 2\phi$ component varies for the different types of magnetic layers we have studied.

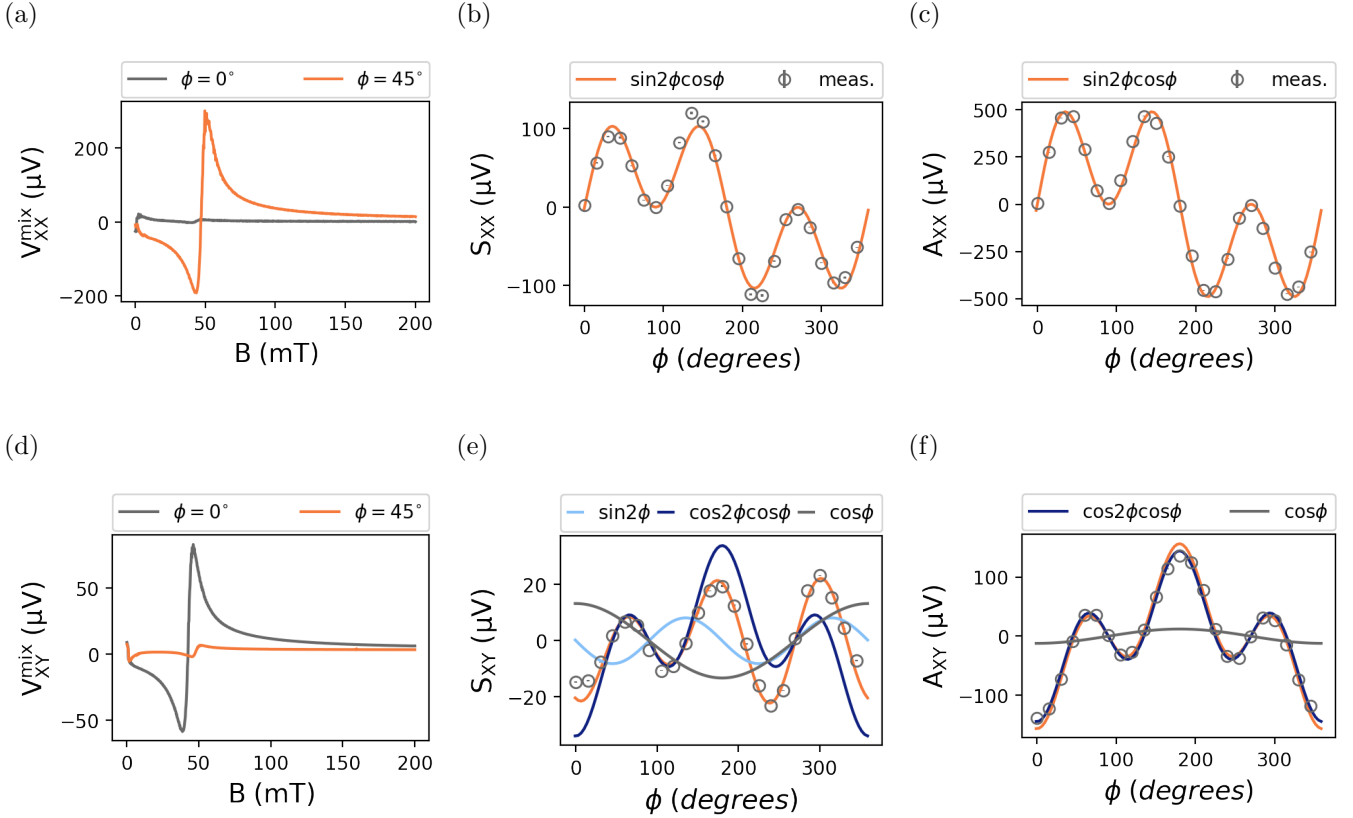

FIG. 4: Results of longitudinal and transverse ST-FMR measurements at room temperature for a Pt(6 nm)/Co$_{90}$Fe$_{10}$(6 nm) sample. (a) Resonance lineshapes for longitudinal ST-FMR. (b,c) Angle dependences of the (b) symmetric and (c) antisymmetric resonance components for longitudinal ST-FMR. (d) Resonance lineshapes for transverse ST-FMR. (e,f) Angle dependences of the (e) symmetric and (f) antisymmetric resonance components for transverse ST-FMR. The lines in (b,c,e,f) show fits to the angular components described in the main text as well as the sums of the fit components.

## IX. LOW-FREQUENCY SECOND HARMONIC HALL MEASUREMENTS

It is widely known (but not explained clearly in the literature) that resonant ST-FMR measurements and non-resonant second-harmonic Hall measurements of spin-orbit torques can differ even for identical layer structures, with results from the low-frequency second harmonic Hall measurements resulting in spin Hall torque efficiencies larger by tens of percent. Since our correction to ST-FMR for the presence of artifact voltages tends to increase the measured spin Hall torque efficiency, we performed second harmonic Hall measurements on our samples to see if the discrepancy

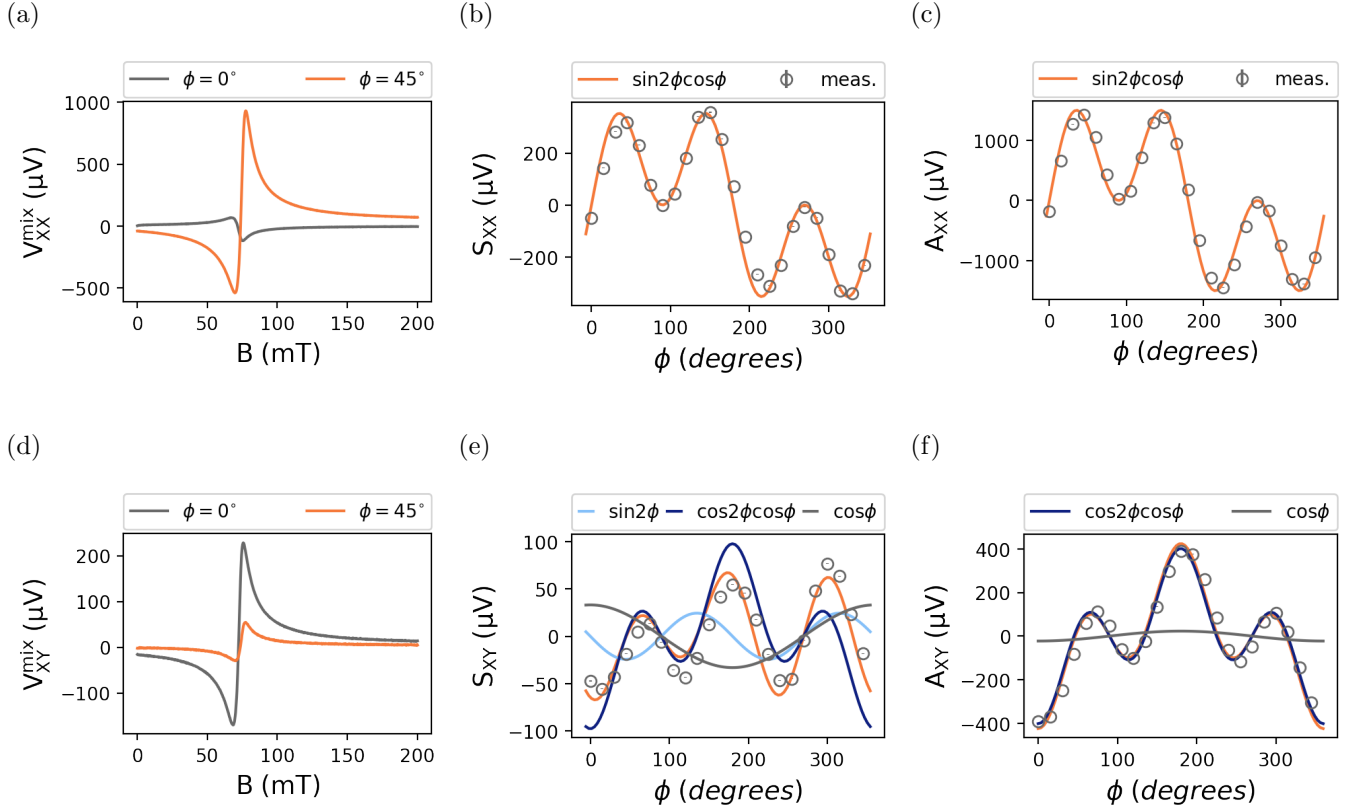

FIG. 5: Results of longitudinal and transverse ST-FMR measurements at room temperature for a Pt(6 nm)/Ni$_{81}$Fe$_{19}$(8 nm) sample. (a) Resonance lineshapes for longitudinal ST-FMR. (b,c) Angle dependences of the (b) symmetric and (c) antisymmetric resonance components for longitudinal ST-FMR. (d) Resonance lineshapes for transverse ST-FMR. (e,f) Angle dependences of the (e) symmetric and (f) antisymmetric resonance components for transverse ST-FMR. The lines in (b,c,e,f) show fits to the angular components described in the main text as well as the sums of the fit components.

seen in previous measurements could be explained.

We carried out these measurements on standard Hall bar shaped devices patterned on the same samples we used for our ST-FMR devices. We performed these measurements on samples with 2 and 8 nm thick CoFeB since these samples represent the extremes of the artifact voltage's effect on ST-FMR measurements. We employed the standard methodology for investigating in-plane magnetized samples [9] which involves measuring the first and second harmonic Hall response as a function of in-plane external field angle. The fitting functions we use for the first and second

|  | CoFeB Avg. | CoFe(6) | Py(8) |
|---|---|---|---|
| $S_{XY}^{2\phi} / S_{XY}$ (%) | 7 | 14.6 | 15.8 |
| $R_{\text{AMR}}/R_{XX}$ (%) | 0.03 | 1 | 0.83 |
| $R_{\text{AHE}}/R_{XX}$ (%) | 0.2 | 0.2 | 0.06 |
| $R_{\text{AMR}}/R_{\text{AHE}}$ | 0.15 | 5.0 | 14 |

TABLE I: Comparison of the relative magnitude of the $\sin 2\phi$ component measured in $S_{XY}$ for different samples, together with a comparison of the size of the anisotropic magnetoresistance and anomalous Hall resistance relative to the overall sample resistance.

harmonics are

$$V_\omega = \frac{IR_{\text{PHE}}}{2} \sin 2\phi \tag{40}$$

$$V_{2\omega} = IR_{\text{PHE}} \frac{\tau_z^0}{\gamma} \frac{\cos 2\phi \cos \phi}{B} - \left( \frac{IR_{\text{AHE}}}{2} \frac{\tau_x^0}{\gamma} \frac{1}{B + \mu_0 M_{\text{eff}}} + V_{\text{ANE}} \right) \cos \phi \tag{41}$$

where $I$ is the current in the bar and $V_{\text{ANE}}$ is the thermal voltage due to the anomalous Nernst effect. The angular dependent data and fits are shown in supplementary Figure 6(a,b) for 8 nm thick CoFeB with a 2000 G applied external field.

The dampinglike torque can be obtained from the magnetic-field dependence of the amplitude of the $\cos \phi$ part of the second harmonic voltage, respectively (supplementary Fig. 6(c). The amplitude of the $\cos \phi$ contribution follows the expected linear trend well. From this, we obtain a dampinglike torque efficiency of $\xi_{\text{DL}} = 0.147 \pm 0.003$ for the 2 nm CoFeB film and $0.145 \pm 0.008$ for the 8 nm film, both roughly 60% higher than for the corrected ST-FMR measurements. The discrepancy in the dampinglike torque efficiency between the two measurement techniques remains, indicating that the artifact voltages that we correct for in this paper cannot explain the difference.

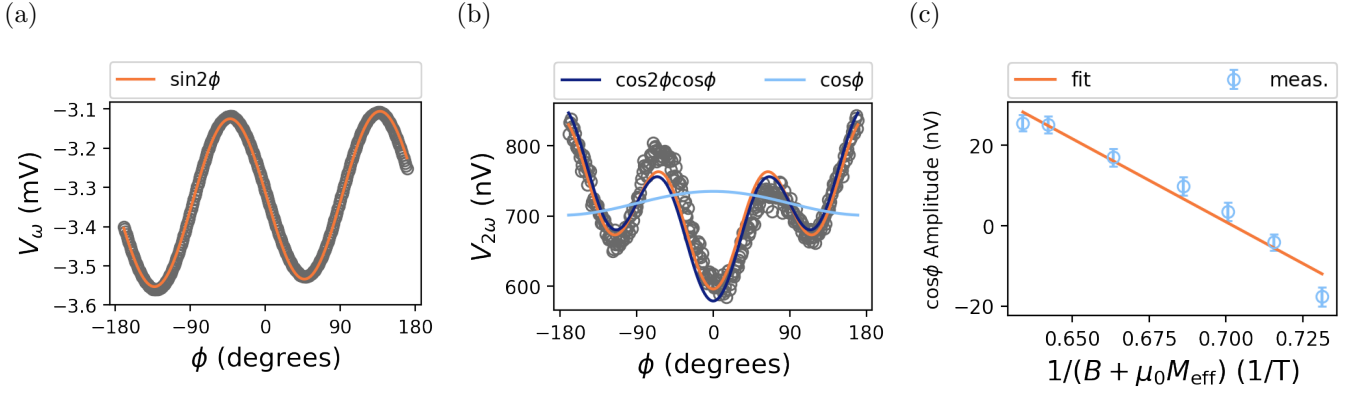

FIG. 6: **(a)** The angular dependence of the first harmonic Hall voltage which allows extraction of the planar Hall resistance. **(b)** The angular dependence of the second harmonic Hall voltage, showing the decomposition into $\cos\phi$ (light blue) and $\cos 2\phi \cos\phi$ (dark blue) components which relate to the dampinglike and fieldlike torques, respectively. **(c)** Field dependence of the $\cos\phi$ component amplitude of the second harmonic Hall voltage. The slope of this line relates to the dampinglike torque and the intercept to the anomalous Nernst voltage. All data shown here are from measurements on the sample with 8 nm of CoFeB.

[1] J. Slonczewski, Journal of Magnetism and Magnetic Materials **159**, L1 (1996).

[2] M. Harder, Z. X. Cao, Y. S. Gui, X. L. Fan, and C.-M. Hu, Phys. Rev. B **84**, 054423 (2011).

[3] N. Vlietstra, J. Shan, V. Castel, B. J. van Wees, and J. Ben Youssef, Phys. Rev. B **87**, 184421 (2013).

[4] Y. Tserkovnyak, A. Brataas, and G. E. W. Bauer, Phys. Rev. Lett. **88**, 117601 (2002).

[5] Y. Tserkovnyak, A. Brataas, and G. E. W. Bauer, Phys. Rev. B **66**, 224403 (2002).

[6] O. Mosendz, V. Vlaminck, J. E. Pearson, F. Y. Fradin, G. E. W. Bauer, S. D. Bader, and A. Hoffmann, Phys. Rev. B **82**, 214403 (2010).

[7] S. Mizukami, Y. Ando, and T. Miyazaki, Japanese Journal of Applied Physics **40**, 580 (2001).

[8] A. Brataas, Y. Tserkovnyak, G. E. W. Bauer, and P. J. Kelly, arXiv:1108.0385 (2011), arXiv:1108.0385 [cond-mat.mes-hall].

[9] C. O. Avci, K. Garello, M. Gabureac, A. Ghosh, A. Fuhrer, S. F. Alvarado, and P. Gambardella, Phys. Rev. B **90**, 224427 (2014).



\end{document}